\documentclass[journal,twocolumn,14pt,twoside]{IEEEtranTCOM}
\usepackage{amsmath}
\usepackage{amsthm,amssymb,amsmath,bm}
\usepackage{subfigure}
\usepackage{amsfonts}
\usepackage{epsfig}
\usepackage{amssymb}
\usepackage{amsmath}
\usepackage[table,xcdraw]{xcolor}
\usepackage{cite}
\usepackage[utf8x]{inputenc}
\usepackage{color,soul}
\usepackage{subfigure}
\usepackage{multirow}
\usepackage{rotating}
\usepackage{graphicx}
\usepackage{tabularx}
\usepackage{array}
\usepackage{color,soul}
\usepackage{yfonts,color}
\usepackage{siunitx}
\usepackage{lettrine}
\usepackage{bm}
\usepackage{graphicx,dblfloatfix}
\usepackage{blindtext}

\usepackage{subfigure}
\usepackage{moreverb}
\usepackage{epsfig}
\usepackage{amsmath,amssymb,amsthm,mathrsfs,amsfonts,dsfont}
\usepackage{adjustbox,lipsum}
\usepackage{amsfonts}
\usepackage{epsfig}
\usepackage{amssymb}
\usepackage{amsmath}
\usepackage{amsthm}
\usepackage{subfigure}
\usepackage{multirow}
\usepackage{rotating}
\usepackage{graphicx}
\usepackage{tabularx}
\usepackage{array}
\usepackage{anyfontsize}
\usepackage{color,soul}
\usepackage{graphicx,dblfloatfix}
\usepackage{epstopdf}
\usepackage{blindtext}
\usepackage{amsmath}
\usepackage{amsthm,amssymb,amsmath,bm}
\usepackage{subfigure}
\usepackage{amsfonts}
\usepackage{epsfig}
\usepackage{amssymb}
\usepackage{amsfonts}
\usepackage{amsmath}
\usepackage{cite}
\usepackage{graphicx}
\usepackage{fancyhdr}
\usepackage{subfigure}
\usepackage[subfigure]{tocloft}
\usepackage[font={small}]{caption}
\usepackage{subfigure}
\usepackage{tabularx}
\usepackage{tcolorbox}
\usepackage{cite}
\usepackage[linesnumbered,ruled,vlined]{algorithm2e}
\SetKwInput{KwInput}{Input}
\SetKwInput{KwOutput}{Output}
\usepackage{amsthm,amssymb,amsmath,bm}
\usepackage{graphicx}
\usepackage{fancyhdr}
\usepackage{subfigure}
\usepackage[subfigure]{tocloft}
\usepackage[font={small}]{caption}
\usepackage{subfigure}
\usepackage{tabularx}

\usepackage{cite}
\allowdisplaybreaks
\usepackage[colorlinks,bookmarksopen,bookmarksnumbered,citecolor=red,urlcolor=red]{hyperref}

\begin{document}
		\author{\IEEEauthorblockN{Mohammad~Reza~Maleki, Mohammad~Robat~Mili, Mohammad Reza Javan,  \IEEEmembership{Senior Member, IEEE}, Nader~Mokari, \IEEEmembership{Senior Member, IEEE}, and Eduard~A.~Jorswieck, \IEEEmembership{Fellow, IEEE}
			\thanks{  
				M.~R.~Maleki, 
				 and N.~Mokari are with the Department of Electrical and Computer Engineering,~Tarbiat Modares University,~Tehran,~Iran, (e-mail: \{M.mohammadreza, nader.mokari\}@modares.ac.ir). M.~R.~Javan is with the Department of Electrical Engineering, Shahrood University of Technology, Iran, (e-mail: javan@shahroodut.ac.ir). E~A.~Jorswieck is with the Institute of Communications Technology, TU Braunschweig, Germany (e-mail: jorswieck@ifn.ing.tu-bs.de). This work was supported by the joint Iran national science foundation (INSF) and German research foundation
				 (DFG) under grant No. 96007867.
	}}}
	\title{ \huge  Multi Agent Reinforcement Learning Trajectory Design and Two-Stage Resource Management in CoMP UAV VLC Networks}
	\maketitle
	\begin{abstract} 
		
		 In this paper, we consider unmanned aerial vehicles (UAVs) equipped with a visible light communication (VLC) access point and coordinated multipoint (CoMP) capability that allows users to connect to more than one UAV. UAVs can move in 3-dimensional (3D) at a constant acceleration, where a central server is responsible for synchronization and cooperation among UAVs. The effect of accelerated motion in UAV is necessary to be considered.  Unlike most existing works, we examine the effects of variable speed on kinetics and radio resource allocations. For the proposed system model, we define two different time frames. In the frame, the acceleration of each UAV is specified, and in each slot, radio resources are allocated.  Our goal is to formulate a multiobjective optimization problem where the total data rate is maximized, and the total communication power consumption is minimized simultaneously. To handle this multiobjective optimization, we first apply the scalarization method and then apply multi-agent deep deterministic policy gradient (MADDPG). We improve this solution method by adding two critic networks together with two-stage resources allocation. Simulation results indicate that the constant acceleration motion of UAVs shows about 8\% better results than conventional motion systems in terms of performance.  
    \end{abstract}
\section{introduction}
\subsection{State of the art}
\lettrine[lines=2]{\color{black}A}{s} a cooperative communication system based on multiple transmission and reception points, coordinated multipoint (CoMP) is consolidated into the long-term evolution-advanced releases \cite{lee2012coordinated} as an adequate method for relieving inter-cell interference. It also enables symbol-level cooperation among  unmanned aerial vehicles (UAVs) and base stations (BS) to enhance communication quality. CoMP technique significantly improves data-rate, and connection availability for \textcolor{black}{cell center} and edge users \cite{liu2019comp, yao2020joint}. \textcolor{black}{Despite the fact that CoMP can mitigate the effects of severe inter-cell interference (ICI), it is considered a key enabling technology for beyond fifth-generation (B5G) and sixth-generation (6G) networks. In order to improve network coverage for the next-generation mobile phone networks, CoMP is used to increase the received signal-to-interference-plus-noise ratio (SINR)} \cite{elhattab2020comp}. \\
UAVs are predicted to play an essential role  B5G and 6G cellular networks \cite{zeng2019accessing}. On the one hand, for improving the communication and service range and enhancing the quality of service (QoS), UAVs with specific purposes such as aerial maneuvers can be linked directly with cellular BSs \cite{zeng2018cellular,shakhatreh2019unmanned , ullah2020cognition, wu2021intelligent}. On the other hand, we can utilize UAVs as aerial wireless BSs  in the sky to implement flexible and on-demand wireless services to mobile users, promoting communication performance and improved coverage \cite{zeng2016wireless, mozaffari2019tutorial, wang2019deployment, li2018uav, zeng2019energy}. Several technical opportunities and challenges are created with the advent of cellular-connected UAVs and wireless transmissions aided by UAVs. As a first consideration, UAVs usually have a strong line-of-sight (LoS) to users. As a result, channel gain and communication quality are improved, but inter-cell interference increases correspondingly. Second, UAVs offer high mobility in the 3-dimensional (3D) environment. \textcolor{black}{Trajectory management becomes more complex due to 3D motion}, but it provides more opportunity for UAV positioning and trajectory control, which can improve communication performance \cite{yao2020joint}. Visible light communication (VLC) is an evolving communication technology with low energy consumption and flexible coverage\cite{bykhovsky2014multiple4}. The VLC network can support a large number of services due to the available bandwidth in unlicensed spectrum, its ubiquitous presence, and low power consumption. Using light-emitting diodes (LEDs) in VLC, the technology offers illumination and communication in scenarios such as search and rescue. It will play an essential role in future generations 
 \cite{chen2017performance6,chen2019downlinkCell-freevlc}. With the emergence of new technologies and applications, machine learning becomes more prevalent in B5G wireless applications in \cite{wang2020deep}.\\
 \textcolor{black}{In this paper, we present a unified framework addressing  these concerns and utilizing different emerging technologies. By designing a complex two time frame system for resource allocation and \textcolor{black}{ trajectory planning}, we provide an approach that solves the multifaceted problem. \textcolor{black}{Considering the constant acceleration makes our work more practical.} We can solve complex problems by combining multiple technologies with novel reinforcement learning (RL) algorithms. We adopt and extend these technologies to maximize the total data rate while users are moving, which is  challenging in VLC networks. However, our paper is different compared to previous works. We design a new trajectory planning with two time frames and utilize novel RL with two-stage actions.} \subsection{Related works}
	We organize the review according to several system models. These categories include UAV-RL, UAV-CoMP, and VLC-RF, respectively, as follows:
\textcolor{black}{\subsubsection{UAV-RL}
QoS constrained energy efficiency function is proposed by the authors in \cite{cui2019multi} as a reward function for providing reliable communication. To deal with dynamics and uncertainty in environments, they form a stochastic game for maximizing the expected rewards and solve the problem by using the multi-agent reinforcement learning (MARL), where each UAV acts as an agent, which seeks the best policy independently and only relies on each agent's local observations. As a way to support the maximum number of offloaded tasks while maintaining the QoS requirements, each \textcolor{black}{mobile edge computing (MEC)} server is assigned unique optimization problems to jointly manage the MEC-mounted macro eNodeBs (MeNBs) to allocate resources to vehicles and make
association decisions in \cite{peng2020multi}. Through the use of the MEC servers as agents, they convert the transformed problems into agents, and then develop the MADDPG algorithm to solve them. In high mobility and highly heterogeneous networks, \cite{tang2020deep} investigates the channel model and develops a novel deep RL-based time division duplex (TDD) configuration method for dynamically allocating radio resources online. A joint scheduling approach between the UAVs' trajectory planning and time resource assignment is implemented to maximize minimum throughput under the constraints of maximum flight speed, peak uplink power, and area of flight in \cite{tang2020minimum}. To maximize the minimum throughput, they propose a multi-agent deep Q-learning (DQL)-based strategy for jointly optimizing the paths and time allocation of the UAVs. Every UAV possesses an independent deep Q-network (DQN) for its own action strategy, while the rest of the UAVs are considered parts of the environment. Specifically, the authors of \cite{zhang2020caching} develop an architecture for delivering content to ground users in a hotspot area using UAV NOMA cellular networks enabling caching. By optimizing the caching placement of a UAV, the scheduling of content requests by users, and the power allocation of NOMA users, they formulate an optimization problem to minimize content delivery delay as a Markov decision process (MDP). They propose an algorithm based on Q-learning that allows the UAV to learn and select action scenarios based on the MDP.}
	\begin{table}
		\footnotesize
		\caption{Notations and Symbols}
		\label{Notations}
		\centering	
		\begin{tabular}{ll}
			\hline Notation \ \ \  \ \     & Description \\
			\hline${m/ M /\mathcal{M}}$ & Index/number/set of users  \\
			${f/ F /\mathcal{F}}$ & Index/number/set  of UAVs \\
			${t/ T /\mathcal{T}}$ & Index/number/set  of time frames \\
			${n/ N /\mathcal{N}}$ & Index/number/set time slots \\
			$\mathbf{w}^{m}$ & Location of each user [\si[per-mode=symbol]{\meter}$\times$\si[per-mode=symbol]{\meter}$\times$\si[per-mode=symbol]{\meter} ] \\
			$\mathbf{v^{m}}$ & Velocity of each user [\si[per-mode=symbol]{\meter\per\second} $\times$\si[per-mode=symbol]{\meter\per\second} $\times$\si[per-mode=symbol]{\meter\per\second} ] \\ 
			$\mathbf{q}^{f}$  & Location of the ${f}_\text{th}$ UAV [$m$$\times$$m$$\times$$m$]\\
			$\mathbf{v}^{f}$  & Velocity of the $f_\text{th}$  UAV [\si[per-mode=symbol]{\meter\per\second}  $\times$\si[per-mode=symbol]{\meter\per\second}  $\times$\si[per-mode=symbol]{\meter\per\second} ]\\
			$\mathbf{a}^{f}$  & Acceleration of the ${f}_\text{th}$ UAV [\si[per-mode=symbol]{\meter\per\second^{2}} $\times$\si[per-mode=symbol]{\meter\per\second^{2}} $\times$\si[per-mode=symbol]{\meter\per\second^{2}}]\\
			$d^{m,f}$ & Distance between the $f_\text{th}$ UAV and the $m_\text{th}$ user  [\si[per-mode=symbol]{\meter}] \\
			$T_{s}(\cdot)$ & Gain of the optical filter \\
			$g(\cdot)$ & Optical concentrator gain at  PD \\
			$\phi$ & Angle of irradiance \\
			$A_{r}$ & Active area of  PD $i$ \\
			$\varpi$ & Order of Lambertian emission \\
			$\psi$ & Incidence angle between  LED and 
			device \\
			$\psi_c$ & Semi-angle field
			of view (FOV) of  PD \\
			\hline
		\end{tabular}
	\end{table}
	\subsubsection{UAV-CoMP} The authors of \cite{gupta2015survey} survey the top issues facing UAV-based wireless communication networks, and they investigate UAV networks with flying drones, the energy efficiency of UAVs, and seamless handover between UAVs and the ground BSs.
\cite{liu2019comp}  studies performance optimization of UAV placement and movement in multi-UAV  CoMP communication, where each UAV forwards its received signals to a central processor for decoding. Through using a trajectory design to exploit the high mobility of the UAV access point, it is possible to enhance the data rate significantly. However this may result in very long delays for users \cite{xie2020common}, making it problematic for delay-sensitive ultra-reliable low latency communications (URLLC) applications. In \cite{yao2020joint}, CoMP  is investigated with 3D trajectory optimization where  multiple suspicious eavesdroppers are considered. It is demonstrated in \cite{ali2018downlink} that maximizing sum rates for heterogeneous networks can lead to a remarkable enhancement of spectral efficiency for joint transmission CoMP-NOMA for a wide range of access distances.
In \cite{HUA2019joint}, the UAV  coverage problem is addressed in order to either maximize coverage region or enhance the QoS. The authors of \cite{kilzi2020analysis} examine the strategies for incorporating a UAV into a two-cell NOMA COMP system so that the BS can be sustained. A novel VLC/UAV framework is designed in \cite{yang2019power} to both communicate and illuminate while also optimizing the locations of UAVs to minimize the total power consumption. \subsubsection{VLC-RF}
In order to enhance communication coverage, hybrid VLC/RF systems emerge, \cite{kashef2016energy}, so that mobile users can achieve higher rates of data transmission via integrated  VLC/RF. A multi-agent reinforcement learning method is used to improve the QoS for the users in \cite{kong2019q}. An RF/VLC aggregated system is discussed by the authors of \cite{ma2020aggregated} in order to maximize energy efficiency. However, they do not assume the user mobility, which is challenging in RF/VLC hybrid networks. By using NOMA-based hybrid VLC-RF with common backhaul, \cite{papanikolaou2020optimal} addresses the problem of optimal resource allocation to maximize achievable data rate. An iterative algorithm is presented to train users on access networks of a hybrid RF/VLC system in \cite{obeed2018joint}. To maximize the total achievable throughput, they formulate an optimization problem to assign power to RF APs and VLC APs.  The symbol error rate of the code domain NOMA-based VLC system is investigated in \cite{dai2018code}, revealing that users exhibit identical error rate performance across locations, while recent works demonstrate that power domain-NOMA is an effective multiple access scheme for VLC systems. \\
In terms of resource allocation, considering the movement of UAVs is  challenging. First, UAV movement with constant speed is practically impossible. In addition, it remarkably deteriorates maneuverability, however, flying at a constant acceleration with increasing maneuverability provides better opportunities for allocating resources and tracking the users. Likewise, 3D motion improves UAV performance by increasing maneuverability. We also provide a minimum data rate for each user to assist users with weaker channels usually located at the edge of the cell to enhance QoS. \textcolor{black}{  In \cite{ren2018performance} and \cite{estrada2019superimposed}, the authors proved that using 
 NOMA in VLC systems results in better performance.}
\subsection{Contribution}
In this paper, we consider the downlink scenario for UAVs that are equipped with VLC APs. Studies show that 3D motion can improve UAV performance in terms of the allocation of radio resources \cite{sun2019optimal,kalantari2016number}. However, we improve UAV motion by designing a two time frame system applying constant acceleration movement. To enhance the resource allocation framework, we utilize constant acceleration motion to cover the weakness of low maneuverability. It also assists UAVs to hastily reach better locations in terms of allocating resources. Increasing in maneuverability manifests itself to have a higher  system complexity. Obviously, solving complex problems requires a better and more robust learning method. We address this challenge using  two time frame allocation method, which are entirely compatible with the proposed system model. In the frames, the constant acceleration is determined, whereas in the slot, radio resources are allocated, and the initial velocities are calculated.
Our contributions  can be summarized  as follows:
\textcolor{black}{
\begin{itemize}
	\item We propose a mathematical model for the movement of UAVs so that they can move at various velocities. It means that all UAVs' accelerations are different in the frames. This model adds more complexity to the problem.
	\item To formulate the problem, we need a new frame structure that includes two time frames for resource allocation. In frames, the constant acceleration of each UAV is determined, whereas in slots, radio resources are allocated. 
	\item Our goal is to maximize data rates and minimize power consumption simultaneously. We formulate our problem in two time frames to fit the equations of motion, and determining the movements of UAVs and users into our problem. We also consider a minimum data rate required for each  user to improve QoS at the edge of the cell.
	\item By considering UAV trajectory and user movement in a complex frame structure, the problem becomes challenging. We present a novel solution method that adapts to the operation of UAVs and the use of two time frames to allocate resources to deal with. A multi-agent-RL-based solution is used where agents interact with a central server. Using this approach increases the speed of convergence and solves the problem better compared with existing methods.
     \item In the simulation section, we provide a comprehensive review of the system model. The located baselines confirm the superior performance of our solution. The impacts of different terms in the objective function are inspected. The influence of various constraints on the objective function is also studied. We additionally examine the system with  and without CoMP, which shows the positive impact of employing CoMP in our system model. Also, the constant acceleration and constant velocity are examined, in which constant acceleration gives better performance than constant velocity.
\end{itemize}}
The rest of this paper is organized as follows: Section \ref{section:System model} describes the system model for our
considered UAV-VLC-enabled CoMP. Section \ref{section:Problem formulation} formulates  UAV resource allocation and movement
optimization problems to maximize data-rate and the minimum total power consumption. We propose our reinforcement learning (RL) approach in Section \ref{Multi Agent Based  Solution}. Section \ref{section:simulation} includes simulation results and at the end, conclusion is in Section \ref{section:Conclusion}. 
	\\ \textbf{Symbol Notations:} Matrix variables and vector are represented by bold
	upper-case and lower-case letters, respectively. $|.|$ stands for the
	absolute value. $\mathcal{S}$ indicates set $\{1,2, \ldots, S\}$ and $|\mathcal{S}|=S$ is the
	cardinality of set $\mathcal{S}$. Transpose is indicated by $\left(\cdot\right)^T$. $\lVert \cdot \rVert$ represents the Euclidean norm, $\lvert \cdot \rvert$ stands for the absolute value. $\mathbb{E}\left\{\cdot\right\}$ is an exception operator. Set of real numbers is represented as $\mathbb{R}$.

\section{System model}
\label{section:System model}
\begin{figure}
	\begin{center}
		\centering
		\includegraphics[width=\linewidth]{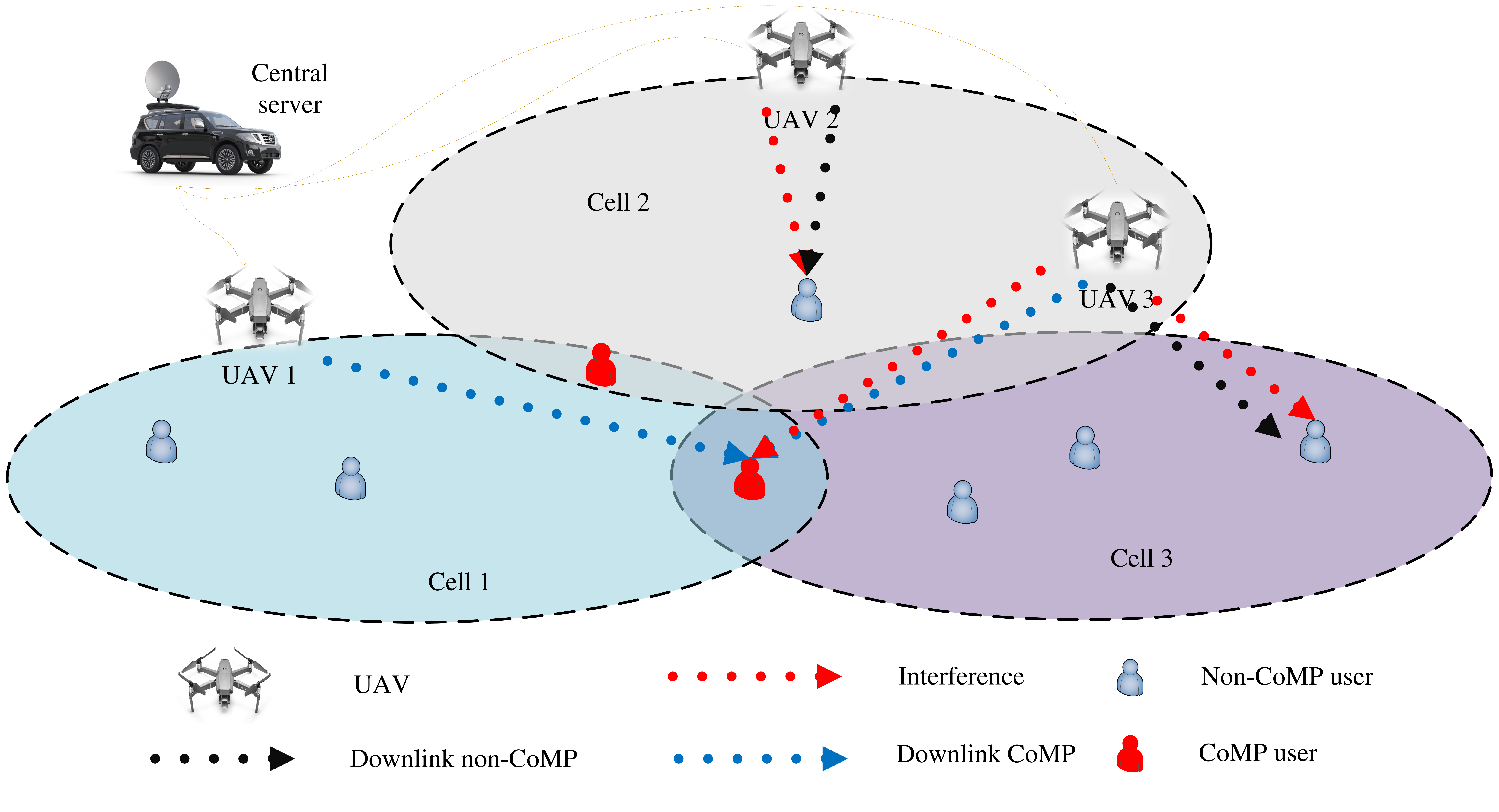}
		\caption{System Modal UAV VLC CoMP}
		\label{sim_opt_eff}
	\end{center}
\end{figure}

\begin{figure}
	\centering
	\includegraphics[width=\linewidth]{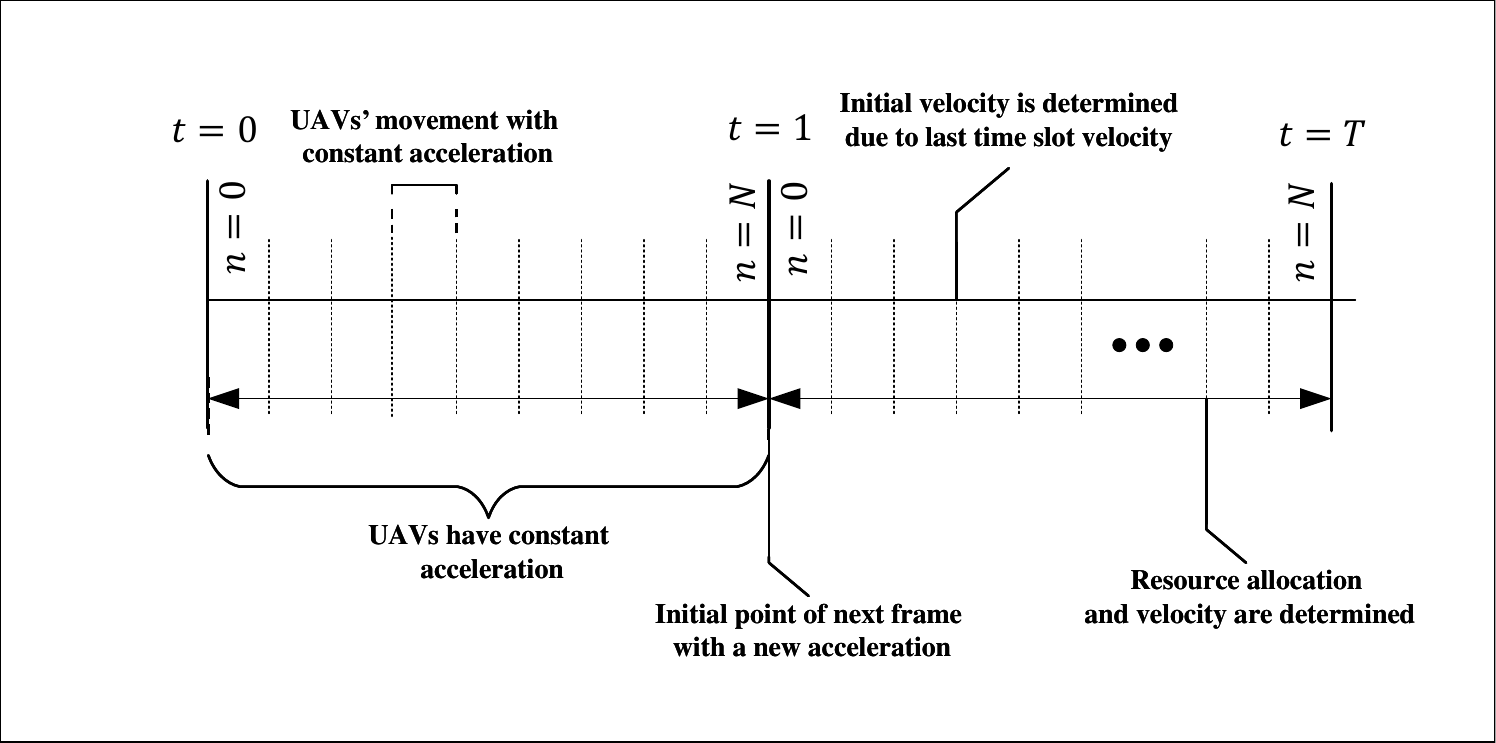}
	\caption{The two frame uses constant acceleration, allocates resources, and calculates the initial velocity to start a slot. }
	\label{fig:time-line}
\end{figure}

In this paper, we consider UAVs mounted with VLC AP, where 
the   UAVs 
 hover  over ground with CoMP system while applying PD-NOMA  to serve both communication and illumination simultaneously.  Users 
 are randomly distributed on ground. As shown in Fig. \ref{sim_opt_eff}, we consider the downlink transmission scenario where  single-antenna UAVs are deployed as aerial BSs. \textcolor{black}{ For ease of exposition, the time horizon, $T$, is equally divided into $N$ time slots with slot duration $\delta$ as shown in Fig. \ref{fig:time-line} where $ \mathcal{T}=\left\{1, \dots, T\right\}$, $ \mathcal{N}=\left\{1, \dots, N\right\}$ are the frame set and the slot set, respectively.} We denote the set of UAVs by $ \mathcal{F}=\left\{1, \dots,F\right\}$, the set of users is indicated by $ \mathcal{M}=\left\{1, \dots,M\right\}$. We consider a 3D Cartesian coordinate system, with location, velocity, and acceleration \textcolor{black}{are  measured in \si[per-mode=symbol]{\meter}, \si[per-mode=symbol]{\meter\per\second}, and \si[per-mode=symbol]{\meter\per\second^{2}}, respectively,} where the horizontal coordinate and velocity of user $m$ are denoted by $\mathbf{w}^{m}[t,n]=$ $\left(x^{m}[t,n], y^{m}[t,n],0 \right)^{\mathrm{T}} \in \mathbb{R}^{3 \times 1}$ and $\{\mathbf{v}^{m}[t,n]\}$  , respectively. We assume that each UAV
flies with the maximum speed constraint $v_{\text {max }}$ and the maximum acceleration constraint $a_{\text {max}}$.  As such, the UAV trajectory, speed, acceleration,  coordinates, velocity, and acceleration of UAV $f$ over time $T$ can be denoted by  $\{\mathbf{q}[t,n]\},\{\mathbf{v}[t,n]\}$, $\{\mathbf{a}[t]\}$, and $\mathbf{q}^{f}[t,n]=$ $\left(x^{f}[t,n], y^{f}[t,n],z^{f}[t,n] \right)^{\mathrm{T}} \in \mathbb{R}^{3 \times 1}$, $\mathbf{v}^{f}[t,n]=$ $\left(v_x^{f}[t,n], v_y^{f}[t,n],v_z^{f}[t,n] \right)^{\mathrm{T}} \in \mathbb{R}^{3 \times 1}$, and  $\mathbf{a}^{f}[t,n]=$ $\left(a_x^{f}[t,n], a_y^{f}[t,n],a_z^{f}[t,n] \right)^{\mathrm{T}} \in \mathbb{R}^{3 \times 1}$,  respectively. 
Assume that the communication channel from UAV to each user is dominated by a line-of-sight (LoS) link. In VLC and UAV networks without loss of generality, the LoS  channel gain of the VLC link between UAV $f$ and
user $m$ can be 
expressed as:
\begin{figure*}[t]
\begin{equation}
	h^{m,f}[t,n]=\left\{\begin{array}{ll}
		\frac{(\tilde{m}+1) A_{r}}{2 \pi\left(d^{m,f}[t,n]\right)^{2}} \cos ^{m}(\phi[t,n])\tilde{F}(\psi[t,n]), &  0 \leq \psi[t,n] \leq \psi_{c}, \\
		0, & \psi_{c} \leq \psi[t,n],
	\end{array}\right.
\end{equation}
\hrule
\end{figure*}
\begin{equation}
	d^{m,f}[t,n]=\left\|\mathbf{q}^{f}[t,n]-\mathbf{w}^{m}[t,n] \right\|,
\end{equation}
and
\begin{equation}
	\tilde{F}(\psi[t,n])= T_{s}(\psi[t,n]) \cos (\psi[t,n]) g(\psi[t,n]),
\end{equation}
where $A_{r}$ is the active area of the photo detector (PD). $d^{m,f}[t,n]$ is the transmission distance from the UAV to the
user and $\psi[t,n]$ denotes the angle of incidence between the UAV and the device,
$\phi[t,n]$ is the angle of irradiance from the UAV to the device. $\tilde{m}$ is the order of Lambertian emission
with $\tilde{m}=-\ln 2 /\left(\ln \cos \phi_{1 / 2}\right)$ where $\phi_{1 / 2}$ is the LED's semi-angle at half power. $T_{s}(\psi[t,n])$ is the
gain of the optical filter and $g(\psi[t,n])$ is the optical concentrator gain at the $\mathrm{PD}$, $g(\psi[t,n])$
can be expressed as: $g(\psi[t,n])=\eta / \sin ^{2} \psi_{c}$ when $0 \geq \psi[t,n] \geq \psi_{c},$ and $g(\psi[t,n])=0$ if $\psi_{c}<\psi[t,n],$ where $\psi_{c}$
and $\eta$ are the semi-angle field of view (FoV) of the PD and the refractive index.\\
 \textcolor{black}{There are two types of users in this system, including CoMP and non-CoMP. A CoMP user is associated with more than one UAV, whereas, a non-CoMP is the same as a traditional network users. The SINR for the $m_\text{th}$  user data, received at user $m$ is expressed as}
\begin{equation}
	\gamma^{m,f}[t,n](m)=\frac{\mu^{2} p^{m,f}[t,n] \rho^{m,f}[t,n]\left| h^{m,f}[t,n]\right|^{2}}{I^{m,f}_\text{ Intra }[t,n]+B_{m,f}\left(\sigma^{m}\right)^{2}},
\end{equation}
where $\mu$ is the PD’s responsibility, $  p^{m,f}[t,n] $ is the allocated power between user $m$ and UAV $f$, $\rho^{m,f}[t,n] $ is the user association variable where $ \rho^{m,f}[t,n] \in \lbrace 0,1 \rbrace$, and $\left|\mathbf h^{m,f}\right| $ is channel coefficient between user $m$ and  UAV $f$.
$I^{m,f}_\text{ Intra }[n] $ is  NOMA interference  and $ (\sigma^{m})^{2}$ is  noise variance. Due to low interference in VLC system, we 
\begin{figure*}
	\centering
	\includegraphics[width=1\linewidth]{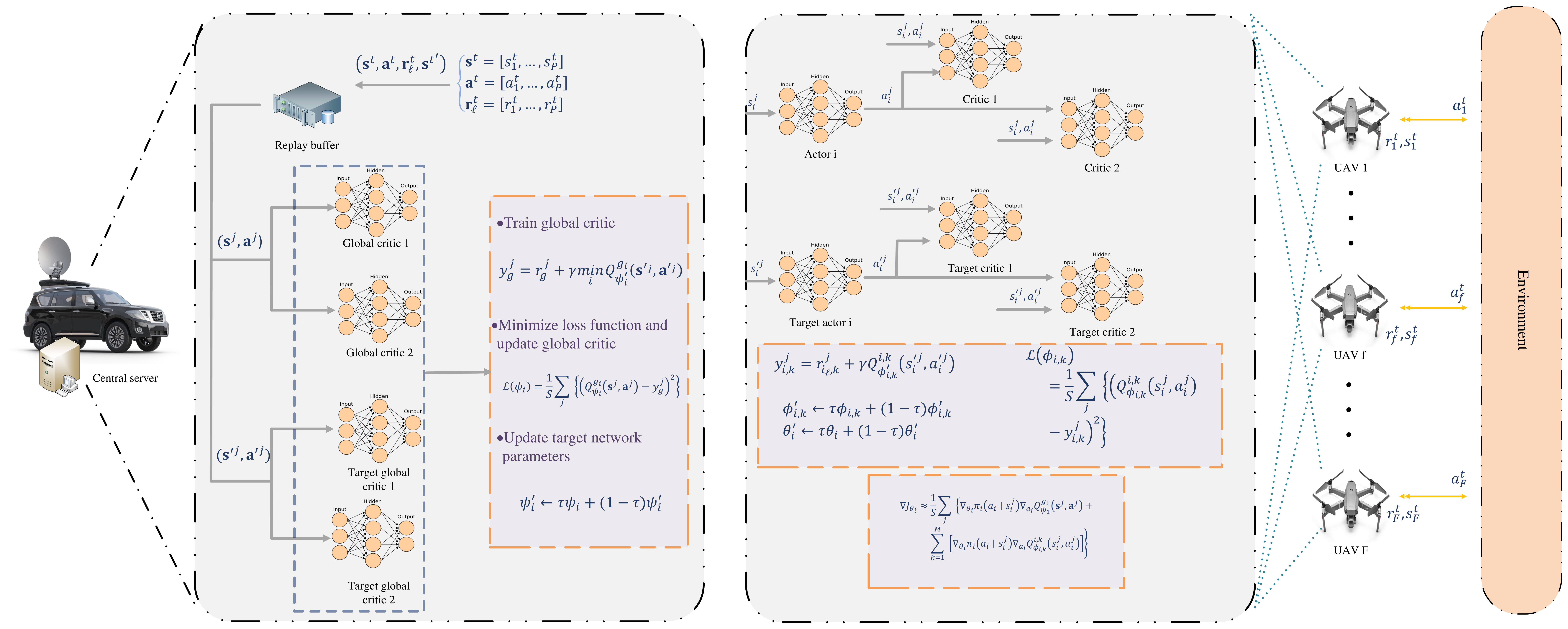}
	\caption{Proposed MADDPG two-stage resource allocation and trajectory planning}
	\label{fig:marl}
\end{figure*} have only NOMA interference in the denominator of SINR. We can indicate $I^{m,f}_\text{ Inter } $ and $I^{m,f}_\text{ Intra } $ due to the principle of NOMA as:
\begin{equation}
	\label{interfence}I^{m,f}_\text{ Intra }[t,n] = \sum_{i\in \mathbf U_m}\rho^{i,f^{\prime}}[t,n] p^{i, f^{\prime}}[t,n]\left| h^{i,f^{\prime}}[t,n]\right|^{2},
\end{equation}
where 
\begin{equation}
	\footnotesize
	\label{ignore}\mathbf{U}^{m}=\left\{b \in M \left\|{h}^{b, f}[t,n]\right|^{2} >\left|{h}^{m, f^{\prime}}[t,n]\right|^{2}\right\},
\end{equation}
In \eqref{interfence}, the user with the best channel ignores other signals and only senses noise as its interference as it is obvious in \eqref{ignore}.\\
We define the total date rate as summation of CoMP and nonCoMP users as:
\begin{equation}
	\tilde{R}[t]= \sum_{m \in \mathcal{M}}\left(R_\text{CoMP}^{m}[t]+\sum_{f \in \mathcal{F}}R_\text{nonCoMP}^{m, f}[t]\right),
\end{equation}
where
\begin{equation}
	R_\text{CoMP}^{ \ m}[t]=\frac{1}{N} \sum_{n=1}^{N} R_\text{CoMP}^{\prime \ m}[t,n],
\end{equation}
\begin{equation}
	R_\text{nonCoMP}^{ \ m,f}[t]=\frac{1}{N} \sum_{n=1}^{N} R_\text{nonCoMP}^{\prime \ m,f}[t,n],
\end{equation}
where
\begin{equation}
	R_\text{CoMP}^{\prime \ m}[t,n] = \nu^{m,f}[t,n]\sum_{f \in \mathcal{F}}\frac {B_{m,f}}{2}
	\log_2 \left(1 + \gamma^{ m,f}[t,n]\right) ,
\end{equation}
\begin{equation}
	R_\text{nonCoMP}^{\prime \ m,f}[t,n] = \left(1-\nu^{m,f}[t,n]\right)\frac {B_{m,f}}{2}
	\log_2 \left(1 + \gamma^{m,f}[t,n]\right),
\end{equation}
where $\nu^{m,f}$ is a binary variable which indicates whether the user is CoMP or not.
The scaling factor $1/2$ is
due to the Hermitian symmetry\cite{zhang2018energy}.\\
\section{Problem formulation}
\label{section:Problem formulation}
In this section, we introduce a multiobjective optimization problem (MOOP) where the data rates of CoMP users and nonCoMP users are maximized, and the total power of these users is minimized simultaneously as follows:
 \begin{subequations}
	\begin{align}
		&\max_{\left\{\mathbf{w}_{m}[t,n],\mathbf{q}[t,n],\mathbf{v}[t,n-1] ,\mathbf{a}[t],\rho^{m,f}[t,n], p^{m,f}[t,n]\right\}}  \tilde{R} [t] \\&  
		\min_{\left\{\mathbf{w}_{m}[t,n],\mathbf{q}[t,n],\mathbf{v}[t,n-1] ,\mathbf{a}[t]\right\}}\sum_{f\in \mathcal F}\sum_{m\in \mathcal{M}} p^{m,f}[t,n], \\&
		s.t.  \label{power-total}\sum_{f\in \mathcal{F}}\sum_{m\in \mathcal{M}}\rho^{m,f}[t,n] p^{m,f}[t,n]\leq p_\text{max}, \ \forall n \in \mathcal{N},\   t \in \mathcal{T}, \\&
		\label{power-per}0 \leq \sum_{m \in M} \rho^{m,f}p^{m,f}[t,n]\leq \tilde p_\text{max}^f,\ \forall f \in \mathcal{F}, \  n \in \mathcal{N}, \  t \in \mathcal{T}, \\&
		\label{rate-comp}R_\text{CoMP}^{m} [t,n] \geq R_\text{min}^{\prime f},\ \forall f \in \mathcal{F}, \ m \in \mathcal{M}, \  n \in \mathcal{N}, \  t \in \mathcal{T}, \\& 
		\label{rate-non}R_\text{nonCoMP}^{m,f}[t,n]\geq R_\text{min}^f ,\ \forall f \in \mathcal{F}, \ m \in \mathcal{M}, \  n \in \mathcal{N}, \  t \in \mathcal{T}, \\& \nonumber
		\label{movement}\mathbf{q}^f[t,n]=\mathbf{q}^f[t,n-1]+\mathbf{v}^f[t,n-1] \delta+\frac{1}{2} \mathbf{a}^f[t] \left(\delta\right)^{2},\\&  \forall f \in \mathcal{F},\ n \in \mathcal{N},\ t \in \mathcal{T},   \\&
		\label{velocity-UAV}\mathbf{v}^f[t,n]=\mathbf{v}^f[t,n-1]+\mathbf{a}^f[t] \delta, \ \forall f \in \mathcal{F},\ n \in \mathcal{N},\ t \in \mathcal{T},   \\&
		\nonumber\label{location-user}\mathbf{w}^m[t,n]=\mathbf{w}^m[t,n-1]+\mathbf{v}^m[t,n-1] \delta,\ \forall m \in \mathcal{M},\\& n \in \mathcal{N},\ t \in \mathcal{T},  \\ &         
		\nonumber \label{UAV-confined}\lVert  (x^{f}[t,n],y^{f}[t,n])-(x_\text{mid},y_\text{mid}) \rVert \leq \hat{r}_\text{Radius}, \\&  0\leq z_{f}[t,n] \ \leq z_\text{max},\ \forall f \in \mathcal{F},  \  n \in \mathcal{N},  t \in \mathcal{T},
		\\&
		\nonumber\label{user-limit}\lVert\mathbf{w}^f[t,n]-(x_\text{mid},y_\text{mid})\rVert\leq \hat{r}_\text{Radius}, \forall f \in \mathcal{F}, \  n \in \mathcal{N}, \\& t \in \mathcal{T},\\& 
		\label{max-velocity-uav}\lVert \mathbf{v}^f[t,n] \rVert \leq v_{\max },  \ \forall f \in \mathcal{F},\ n \in \mathcal{N},  t \in \mathcal{T},  \\&
		\label{max-acc-uav}\lVert\mathbf{a}^f[t]\lVert \leq a_{\max }, \ \forall f \in \mathcal{F},\ t \in \mathcal{T}, \\&
		\label{movement2}\lVert\mathbf{v^m}[n,t]\lVert \leq v^{\prime}_{\max },  \ \forall  m \in \mathcal{M},\ n \in \mathcal{N},  \ t \in \mathcal{T}, \\&
		\label{user-in-cell}\sum_{m\in \mathcal{M}}\rho_{m,f}[t,n]\leq J_K, \   \forall f \in \mathcal{F},\  n \in \mathcal{N}, \ t \in \mathcal{T},  \\&
		\nonumber\gamma^{m,f}[t,n](i)-\gamma^{m,f}[t,n](m)\geq 0,  \\&
		\label{NOMA}, 
		\nonumber 	|h^{i,f}[t,n]|^2 > |h^{m,f}[t,n]|^2 \\&
		\forall  \ f\in \mathcal{F}, \ m, \ i \in \mathcal{M},\  n \in \mathcal{N}, \ t \in \mathcal{T},  \\&
		\label{assianemt}\rho^{m,f}[t,n]\in \lbrace 0,1\rbrace, \ \forall f \in \mathcal{F}, \ m \in \mathcal{M},  n \in \mathcal{N}, \ t \in \mathcal{T}, \\&
		\label{CoMP}\nu^{m,f}[t,n] \in \lbrace 0,1\rbrace, \ \forall f \in \mathcal{F}, \ m \in \mathcal{M},  n \in \mathcal{N}, \ t \in \mathcal{T},
	\end{align}
\label{problem}
\end{subequations}
 \eqref{power-total} demonstrates the maximum power of each UAV that can transmit, \eqref{power-per} is UAVs power constrained between the maximum and zero. \eqref{rate-comp} and \eqref{rate-non} are the minimum data rate constraint which all UAVs need to satisfy.   \eqref{movement}-\eqref{movement2} are movement constraints. \textcolor{black}{ The equations of motion  of each UAV  is shown in \eqref{movement}, \eqref{velocity-UAV} presents the velocity equation  of UAVs, the location of users is obtained from equation \eqref{location-user}. \eqref{UAV-confined} and \eqref{user-limit} illustrate the spatial constraint of the simulation for UAVs and users. \eqref{max-velocity-uav}, \eqref{max-acc-uav}, and \eqref{movement2} indicate the maximum UAV speed, the maximum UAV acceleration, and the maximum speed of each user, respectively. \eqref{user-in-cell} demonstrates the number of users in the service  of each UAV. \eqref{NOMA} are the NOMA constraints which shows the $m_\text{th}$ user SINR at the $i_\text{th}$ user must be bigger than at the $m_\text{th}$ user.  \eqref{assianemt} is the assignment index that specifies assignment of user to UAV and as  a binary variable. \eqref{CoMP} is a binary variable which indicates whether the user is CoMP or not.} \textcolor{black}{
 	\eqref{problem} is a non-convex NP-hard problem that conventional mathematical methods cannot solve. To handle the above problem, in the following, we present the MADDPG approach, a machine learning method suitable for complex problems.}
\section{Multi Agent Based  Solution }
\label{Multi Agent Based  Solution}
\textcolor{black}{Due to the complexity of the problem, we are not able to solve the problem by classical programming methods, so we move to use RL methods. Single-agent methods face problems in estimating and overloading information. Conventional multi-agents methods cannot obtain better results than single-agent methods in small environments due to  utterly independent functionality.
}
In this section, we  form our environment, agents, and the relevant interaction  among the agents, which is states, actions, and reward. This section ends up with formulating our multi-agent approach. 
\subsection{Environment}
The purpose of each agent, in a multi-agent  environment, is to maximize its policy function, which can be shown as:
\begin{equation}
\label{reward-expection}	\max _{\pi^{f}} \mathcal{J}^{f}\left(\pi^{f}\right), \quad f \in \mathcal{F}, \quad \pi^{f} \in \Pi^{f},
\end{equation}
where  $\mathcal{J}^{f}\left(\pi^{f}\right)=\mathbb{E}\left[\sum_{t=0}^{\infty} \gamma^{t} \tilde{r}^{f}_{t+1} \mid s^{f}_{0}\right]$ is our conditional expectation, $\pi_{f}$ is the policy of UAV 
$f$, and the set $\Pi_{j}$ contains all feasible policies that are available for $\mathrm{UAV}$ $f$. Information from the aforementioned discussion about each $\mathrm{UAV}$ as an agent interacts with the UAV network environment and takes action relevant to its policy. The goal is to solve the optimization problem \eqref{problem}. The agents seek to improve their reward \eqref{reward-expection}, which is related to the objective function. At each time $n$, the UAV immediately takes action, $a_{t,n}$ after observing environment state, $s_{t,n}$. The environment transfers to a new state $s_{t,n+1}$ in the transition step and UAV obtains a reward related to its action. Next we describe the state space $\mathcal{S}$, action space $\mathcal{A}$, and the reward function $r_{t,n}$, in our system model: 
\begin{itemize}
	\item State Space: The state of each agent includes all immediate channel gains at time $n$ of users $h^{m,f}[t,n]$ for all $m \in \mathcal{M} \& f \in \mathcal{F}$, all UAV previous velocity and location $\mathbf{q}[t,n-1],\mathbf{v}[t,n-1]$   and users previous location  $\mathbf{w}_{m,k}[t,n-1]$, all interferences involved in the $n-1$ slot, $I^{m,f^{\prime}}_\text{ Intra }[t,n-1]$ and $I^{m,f}_\text{ Inter }[t,n-1]$.
	\begin{align}
		\mathbf{s}_{t,n}^{f}=[&h^{m,f}[t,n],\mathbf{q}[t,n-1],\mathbf{v}[t,n-1], \mathbf{w}_{m,k}[t,n-1],\\ \nonumber
		 &I^{m,f^{\prime}}_\text{ Intra }[t,n-1], I^{m,f}_\text{ Inter }[t,n-1]], \quad f \in \mathcal{F}.
	\end{align}
	\item Action Space: In each time step UAVs (agents) act as  $a_{t,n}^f = \{\rho^{f}_{t,n}, \mathbf{p}^{f}_{t,n}, \nu^{m,f}_{t,n}, \mathbf{a}^{f}_{t} \}$, as we mentioned above, $\rho^{f}_{t,n}$ is our assignment variable,  $\mathbf{p}^{f}_{t,n}$ is the power that is allocated to the users, $\nu^{m,f}_{t,n}$ is CoMP indicator that shows the user is either CoMP or not, and $\mathbf{a}^{f}_{t}$ is the acceleration matrix. Our variables contain both integer and continuous variables. It suggests to adopt policy gradients to solve the problem. Making use of this capability allows us to come up with better solutions.
	\item Reward function: The reward is one of the most important parts of RL because the major driver of RL is the reward. It is crucial to formulate a function accurately that can both represent the objective function and \textcolor{black}{faster and more stable convergence.} First we form our rewards then discuss it. In this system we employ two types of reward, one type per agent and another one is for  global critic network. Agents' goals are to maximize their rates while minimizing  power consumption. \textcolor{black}{The global reward decreases the total interference among the users} as mentioned in \cite{mmad}. All agents have commitment only to their goal and it is not necessary to know about other agents policy, since they share global critic. The goal of the global critic is to connect all the agents together to aid faster convergence  with the support of the reward. To maintain the stable convergence the central server connects the entire system and receives all the states and criticizes the actions of agents. Our global and per agent reward are defined as:
	\begin{equation}
	\label{reward-objective}	\begin{aligned}
			\tilde{r}_{\ell}^{f}=&\alpha\frac{\sum_{m\in \mathcal{M}} \left(R_\text{CoMP}^{m,f} (\rho,p)
				+R_\text{nonCoMP}^{m,f}(\rho,p)\right)}{\frac{B_{m,f}}{2}\log_2\left(1+\frac{\tilde{p}_\text{max}^{f}}{B_{m,f}\left(\sigma^{m}\right)^2}\right)}
			-\\
			&\left(1-\alpha\right)\frac{\sum_{m\in \mathcal{M}} \rho^{m,f}p^{m,f}}{\tilde{p}_\text{max}^{f}},
		\end{aligned}
	\end{equation}
and global reward
\begin{equation}
	\begin{aligned}
		\tilde{r}_{G} = -\left(I^{m,f^{\prime}}_\text{ Intra } + I^{m,f}_\text{ Inter }\right).
	\end{aligned}
\end{equation}
\textcolor{black}{We use linear scalarization to transform our multi-objective problem to single objective using weight factor $\alpha  \in \left(0,1\right)$} \cite{aydin2017energy}.
\end{itemize} 
\subsection{MADDPG}
We assume $\boldsymbol{\pi}$ as a set for all UAVs policies UAV $f$ policy is $\pi^{f}$ $\left(\pi^{f} \in \boldsymbol{\pi} = \{\pi^{1}, \ldots,\pi^{F}\}\right)$ with parameters $\tilde{\theta}^{f}$, $\tilde{Q}^f$ for UAV critic (Q-function) with parameter $\tilde{\phi}^{f}$ and $\tilde{Q}_G$ for global Q-function with parameter $\tilde{\psi}$. Now, we can form our neural network, after that, we can discuss gradient policy. We consider $L^{f}_{\pi}$, $L^{f}_{q}$, and $L_G$ as the number of layers in neural network for each agent action and critic and global critic, respectively. According to the above information, we can develop our neural network in this way $\tilde{\Theta}_{i}=\left(W_{\pi}^{(1)}, \ldots, W_{\pi}^{\left(L_{\pi}\right)}\right)$, $\tilde{\Phi}_{q}=\left(W_{q}^{(1)}, \ldots, W_{q}^{\left(L_{q}\right)}\right)$, and  $\tilde{\Psi}_{G}=\left(W_{G}^{(1)}, \ldots, W_{G}^{\left(L_{G}\right)}\right)$ as actor, UAV critic and global critic. According to the above, the gradient policy is
\begin{equation}
	\nabla_{\theta^{f}} \mathcal{J}^{f}=\mathbb{E}\left[\left.\nabla_{\tilde{\theta}^{f}} \pi^{f}\left(a^{f} \mid s^{f}\right) \nabla^{a^{f}} Q^{f}_{\pi}(\mathbf{s}, \mathbf{a})\right|_{a^{f}=\pi^{f}\left(s^{f}\right)}\right],
\end{equation}
where $\boldsymbol{a} = \left(a^{1},\ldots, a^{F}\right)$ is all actions that are taken by each UAV with observation $\boldsymbol{s} = \left(s^1, \ldots, s^F\right)$. We integrate all actions and states in $Q^{f}_{\pi}(\mathbf{s}, \mathbf{a})$ as inputs to approximate Q-function for UAV $f$. Here we utilize two critic networks, and  we reformulate the policy gradient follows:
\begin{equation}
	\begin{aligned}
		\label{17}\nabla_{\theta^{f}} \mathcal{J}^{f}=& \underbrace{\mathbb{E}_{\mathbf{s}, \mathbf{a} \sim \mathcal{D}}\left[\nabla_{\theta^{f}} \pi^{f}\left(a^{f} \mid s^{f}\right) \nabla_{a^{f}} Q^{\psi}_{G}(\mathbf{s}, \mathbf{a})\right]}_{\text {Global Critic }}+	\\
		& \underbrace{\mathbb{E}_{s^{f}, a^{f} \sim \mathcal{D}}\left[\nabla_{\theta^{f}} \pi^{f}\left(a^{f} \mid s^{f}\right) \nabla_{a^{f}} Q_{\phi^{f}}^{f}\left(s^{f}, a^{f}\right)\right]}_{\text {UAV critic}},
	\end{aligned}
\end{equation} as shown above $a^{f} = \pi^{f}(s^{f})$ are actions which UAV $f$ takes with observation $s^{f}$, respect to policy $\pi^{f}$.
 In \eqref{17}, we have two terms, the first one shows global critic which receives actions and states of all UAVs and it estimates global Q-function using global reward $\tilde{r}_{G}$ and other term shows critic of UAV which receives only itself actions and states. For updating the loss function of global critic, we use
\begin{equation}
	\mathcal{L}(\psi)=\mathbb{E}_{\mathbf{s}, \mathbf{a}, \mathbf{r}, \mathbf{s}^{\prime}}\left[\left(Q^{\psi}_{G}(\mathbf{s}, \mathbf{a})-y_{G}\right)^{2}\right],
\end{equation} 
where $y_{G}$ is a target value of estimation and
\begin{equation}
	y_{G}=\tilde{r}_{G}+\left.\gamma Q^{\psi^{\prime}}_{G}\left(\mathbf{s}^{\prime}, \mathbf{a}^{\prime}\right)\right|_{a^{\prime 
		f}=\pi^{\prime f}\left(s^{\prime f}\right)},
\end{equation}
where our target policy is $\boldsymbol{\pi^{\prime}} = \{\pi^{\prime 1},\ldots, \pi^{\prime F}\}$. We parameterize it with $\boldsymbol{\theta}^{\prime} = \{\theta^{\prime 1}, \ldots, \theta^{\prime F}\}$, and the UAV loss function and its target update as follows:
 \begin{equation}
 	\mathcal{L}^{f}\left(\phi^{f}\right)=\mathbb{E}_{\mathbf{s}^{f}, \mathbf{a}^{f}, \mathbf{r}^{f}, \mathbf{s}^{\prime f}}\left[\left(Q_{\phi^{f}}^{f}\left(s^{f}, a^{f}\right)-y^{f}\right)^{2}\right],
 	\label{loss-update}
 \end{equation}
and $y^{f}$:
\begin{equation}
	y^{f}=r^{f}+\left.\gamma Q_{\phi^{\prime f}}^{f}\left(s^{\prime f}, a^{\prime f}\right)\right|_{a^{\prime f}=\pi^{\prime f}\left(s^{\prime f }\right)}.
	\label{loss-parameter}
\end{equation}
Although this framework is able to reach quite good performance and converge in moderated steps, it still suffer from overwhelmed estimation and its loss in approximations deteriorates framework performance due to sub-optimal policy in Q-function. With results in \cite{mmad} and \cite{van2017hybrid},  we swap global critic with twin delayed deterministic policy gradient in \eqref{172}, then our new gradient policy is
\begin{equation}
	\begin{aligned}
		\label{172}\nabla_{\theta^{f}} \mathcal{J}^{f}=& \underbrace{\mathbb{E}_{\mathbf{s}, \mathbf{a} \sim \mathcal{D}}\left[\nabla_{\theta^{f}} \pi^{f}\left(a^{f} \mid s^{f}\right) \nabla_{a^{f}} Q^{\psi_i}_{G_i}(\mathbf{s}, \mathbf{a})\right]}_{\text {TD3 Global Critic }}+	\\
		& \underbrace{\mathbb{E}_{s^{f}, a^{f} \sim \mathcal{D}}\left[\nabla_{\theta^{f}} \pi^{f}\left(a^{f} \mid s^{f}\right) \nabla_{a^{f}} Q_{\phi^{f}}^{f}\left(s^{f}, a^{f}\right)\right]}_{\text {UAV critic}},
	\end{aligned}
\end{equation}
and loss function
\begin{equation}
	\mathcal{L}(\psi_i)=\mathbb{E}_{\mathbf{s}, \mathbf{a}, \mathbf{r}, \mathbf{s}^{\prime}}\left[\left(Q^{\psi_i}_{G_i}(\mathbf{s}, \mathbf{a})-y_{G}\right)^{2}\right],
\end{equation} 
and $y_G$ is updated as
\begin{equation}
	y^{f}=r^{f}+\left.\gamma(1-\tilde{d})\min_{i=1,2} Q_{\phi_i^{\prime f_i}}^{f_i}\left(s^{\prime f}, a^{\prime f}\right)\right|_{a^{\prime f}=\pi^{\prime f}\left(s^{\prime f }\right)}.
\end{equation}
By \eqref{loss-update} and \eqref{loss-parameter}, the agents critic network  are updated. Our proposed  algorithm is summarized in Algorithm \ref{algorithm} and with intuitive illustrated in Fig. \ref{fig:marl}. Now, we are going to discuss the reasons for choosing the rewards of system. Our first reward is directly related to the objective of the problem, but is calculated independently for each UAV. Each UAV strives to get as close to its maximum and best as possible by maximizing its reward.  \textcolor{black}{Due to the neural network's weakness in calculating interference,} which plays a very important role in the management of radio resources, our global reward maximizes the symmetry of the total system interference. We apply TD3 because it trains global with  with two extra networks, $\tilde{d}$ is delay hyper-parameter.
\begin{algorithm}[ht]
	\small
	\DontPrintSemicolon
	\renewcommand{\arraystretch}{0.9}
	\caption{Two time frame modified MADDPG }
	\label{algorithm}
	Initiate environment, generate UAVs and users \;  
	$\mathbf{Inputs}$: Enter number of $a_t$, $s_t$ agents and users \;
	Initialize all, global critic networks $Q^{f}_{\phi_1}$ and $Q^{f}_{\phi_1}$, target global  critic networks $Q^{\prime f}_{\phi_1}$ and $Q^{\prime f}_{\phi_1}$ and agents policy and critic networks.\;
	\For{t=1 to T}
	{
		\For{n = 1 : $N$ do}
		{\For{each agnet $f$ do}
			{ Observe state $s^{f}_t$ and take action $a^{f}_t$ }
			$\mathbf{s}_{t}=\left[s^{1}_{t}, \ldots, s^{F}_{t}\right], \quad \mathbf{a}_{t}=\left[a_{t}^{}, \ldots, a_{t}^{F}\right]$.\;
			Receive global and local rewards, $\tilde{r}_{G,t}$ and $\tilde{\mathbf{r}}_{t}^{f}$\; Store $\left(\mathrm{s}_{t}, \mathbf{a}_{t}, \tilde{\mathbf{r}}_{t}^{f}, \tilde{r}_{G,t}, \mathbf{s}_{t+1}\right)$ in replay buffer $\mathcal{D}$
		}
		Sample minibatch of size $\mathrm{S},\left(\mathrm{s}^{j}, \mathbf{a}^{j}, \mathbf{r}_{g}^{j}, \mathbf{r}_{\ell}^{j}, \mathbf{s}^{\prime} j\right)$, from
		replay buffer $\mathcal{D}$\; Set $y_{g}^{j}=r_{g}^{j}+\gamma \min _{i} Q_{\psi_{i}^{\prime}}^{g_{i}}\left(\mathbf{s}^{\prime}, \mathbf{a}^{\prime} j\right)$\;
		Update global critics by minimizing the loss:\;
		\begin{equation*}
			\mathcal{L}\left(\psi_{i}\right)=\frac{1}{S} \sum_{j}\left\{\left(Q_{\psi_{i}}^{g_{i}}\left(\mathrm{~s}^{j}, \mathbf{a}^{j}\right)-y_{g}^{j}\right)^{2}\right\}.
		\end{equation*}
		\;
		Update target parameters: $\psi_{i}^{\prime} \leftarrow \tau \psi_{i}+(1-\tau) \psi_{i}^{\prime}$\;
		\If{episode mod $d$ then}{Train actor and critic nerwork\;
			\For{for each agent f do}{  episode mod $d$ then
		$\mid \begin{aligned}&\mathcal{L}\left(\phi_{i}\right)=\frac{1}{S} \sum_{j}\left\{\left(Q_{\phi_{i}}^{i}\left(s_{i}^{j}, a_{i}^{j}\right)-y_{i}^{j}\right)^{2}\right\}. \\&\text { Update local actors: } \\&\nabla J_{\theta_{i}} \approx \frac{1}{S} \sum_{j}\left\{\nabla_{\theta_{i}} \pi_{i}\left(a_{i} \mid s_{i}^{j}\right) \nabla_{a_{i}} Q_{\psi_{1}}^{g_{1}}\left(\mathbf{s}^{j}, \mathbf{a}^{j}\right).\right.\end{aligned}$
		$\left.\nabla_{\theta_{\imath}} \pi_{i}\left(a_{i} \mid s_{i}^{j}\right) \nabla_{a_{i}} Q_{\phi_{i}}^{i}\left(s_{i}^{j}, a_{i}^{j}\right)\right\}$\;
		Update target networks parameters:
		$\left[\begin{array}{l} \theta_{i}^{\prime} \leftarrow \tau \theta_{i}+(1-\tau) \theta_{i}^{\prime}, \\ \phi_{i}^{\prime} \leftarrow \tau \phi_{i}+(1-\tau) \phi_{i}^{\prime}.\end{array}\right.$
	}}}
\end{algorithm}
	\begin{table}
	\caption{Simulation parameters}
	\label{simulation}
	\centering	
	\begin{tabular}{ll}
		\hline UAV-VLC CoMP-enabled environment parameters & Value \\
		\hline  FoV $\Psi_{c}$ & $60^{\circ}$\cite{zhang2018energy} \\
		 Detector area of PD $A_{\mathrm{PD}}$ & $1 \mathrm{~cm}^{2}$\cite{zhang2018energy} \\
		$p_{\text {max }}$ & 36 W \\
		$\tilde{p}^{f}_{\text {max }}$ & 12 W \\
		 Number of user & 9 \\
		 Number of UAV & 3 \\
		 Half power angle,$\theta_{1 / 2}$ & $30^{\circ}$\cite{zhang2018energy} \\
		 The order of the Lambertian emission $\varpi$ & 1 \\
		 PD responsivity & $0.53 \mathrm{~A} / \mathrm{W}$\cite{zhang2018energy} \\
		 Sigma-noise & $10 \mathrm{e}-12$ \\
		 $R^{m,f}_\text{min}$ & $0.1 (kbps)$ \cite{savi2019impact} \\
		  $R^{\prime m,f}_\text{min}$ & $0.1 (kbps)$ \cite{savi2019impact} \\
		 $J_K$ & 3 \\
		 $x_\text{mid}$ & 25 m \\
		 $x_\text{mid}$ & 25 m \\
		 $z_\text{max}$ & 100 m \\
		 $a_\text{max}$ & 2 $\sqrt[2]{3}$ \si[per-mode=symbol]{\meter\per\second^{2}} \\
		 $v_\text{max}$ & 10 $\sqrt[2]{3}$ \si[per-mode=symbol]{\meter\per\second} \cite{liu2020machine,tang2020deep} \\
		 $v^{\prime}_\text{max}$ & 5 $\sqrt[2]{2}$ \si[per-mode=symbol]{\meter\per\second} \cite{tang2020deep} \\
		 T & 500 s $\times$ 10e-1\\
		 N &  100 ms\\
		 $\delta$ & 1 ms
	\end{tabular}
\begin{tabular}{ll}
	\hline Neural networks hyper-parameters & Value \\
	\hline Experience replay buffer size & 50000 \\
	Mini batch size & 64 \\
	Number/size of local actor networks hidden layers & $2 / 1024,512$ \\
	Number/size of local critic networks hidden layers & $2 / 512,256$ \\
	Number/size of global critic hidden layers & $3 / 1024,512,256$ \\
	Critic/Actor networks learning rate & $0.001 / 0.0001$ \\
	Discount factor & $0.99$ \\
	Target networks soft update parameter, $\tau$ & $0.0005$ \\
	Number of episodes & 500 \\
	Number of iterations per episode & 100 \\
	\hline
\end{tabular}
\end{table}
\begin{figure}
	\centering
	\includegraphics[width=1\linewidth]{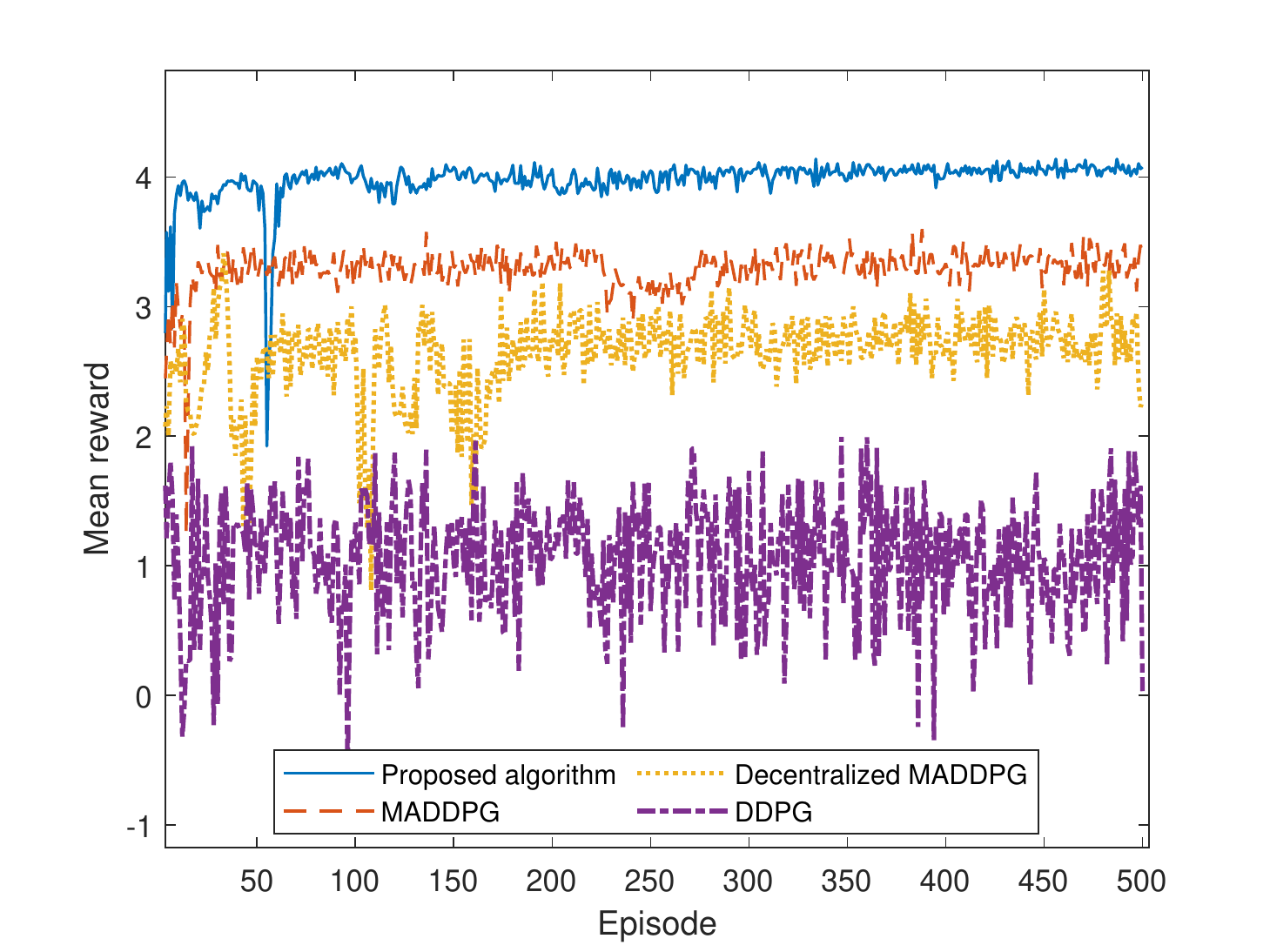}
	\caption{Comparing our approach with MADDPG dec-MADDPG, DDPG.}
	\label{fig:baseline}
\end{figure}
\subsection{Computational Complexity}
In this section we investigate the computational complexity of the algorithm. We divide it into  two sorts and address them in their dedicated section. Complexity is related to $a)$ number of trainable variables and $b)$ total neural network applied to network.
\paragraph[short title]{ Number of trainable variable} The input of the Q-function  includes all actions and states of agents. a In MADDPG, it is similar, but all action and states are specific for each agent. If we assume $\omega$ and $\tilde{\alpha}$ as observation and action space, number of trainable variable of MADDPG will be $\mathcal{O}\left(c^{2}(\omega+\tilde{\alpha})\right)$, where $c$ indicates the number of agents. But in the utilized algorithm, we apply global critic which is being shared among agents. Hence complexity  represented as $\mathcal{O}\left(c(\omega+\tilde{\alpha})\right)$, on the other hand, for UAV critic, it only observes its own agent,  states and action space and fed by them. For each UAV critic we have $\mathcal{O}\left(\omega+\tilde{\alpha}\right)$ as parameter space.
\paragraph[short title]{Number of the nodes in the neural network} \textcolor{black}{In conventional MADDPG the total number of network is $2 \times\left(c\left(\underline{1}_{Q}+\underline{1}_{A}\right)\right)$.This multiplying by 2 is because of one extra network as target network.  $\underline{1}_{Q}$ and $ \underline{1}_{A}$ are the numbers of critic and actor networks, respectively, and $c$ indicates the number of agents. In our proposed algorithm, we applied TD3 and due to 2 critics in its global network, we can demonstrate the total number of the network as  $2 \times\left(c\left(\underline{1}_{Q}+\underline{1}_{A}\right)+\underline{2}_{Q_G}\right)$, where $\underline{2}_{Q_G}$ indicates 2 critic networks in our global critic.}
\begin{figure*}[ht!]
	\centering
	\includegraphics[width=1\linewidth, height=0.4\textheight]{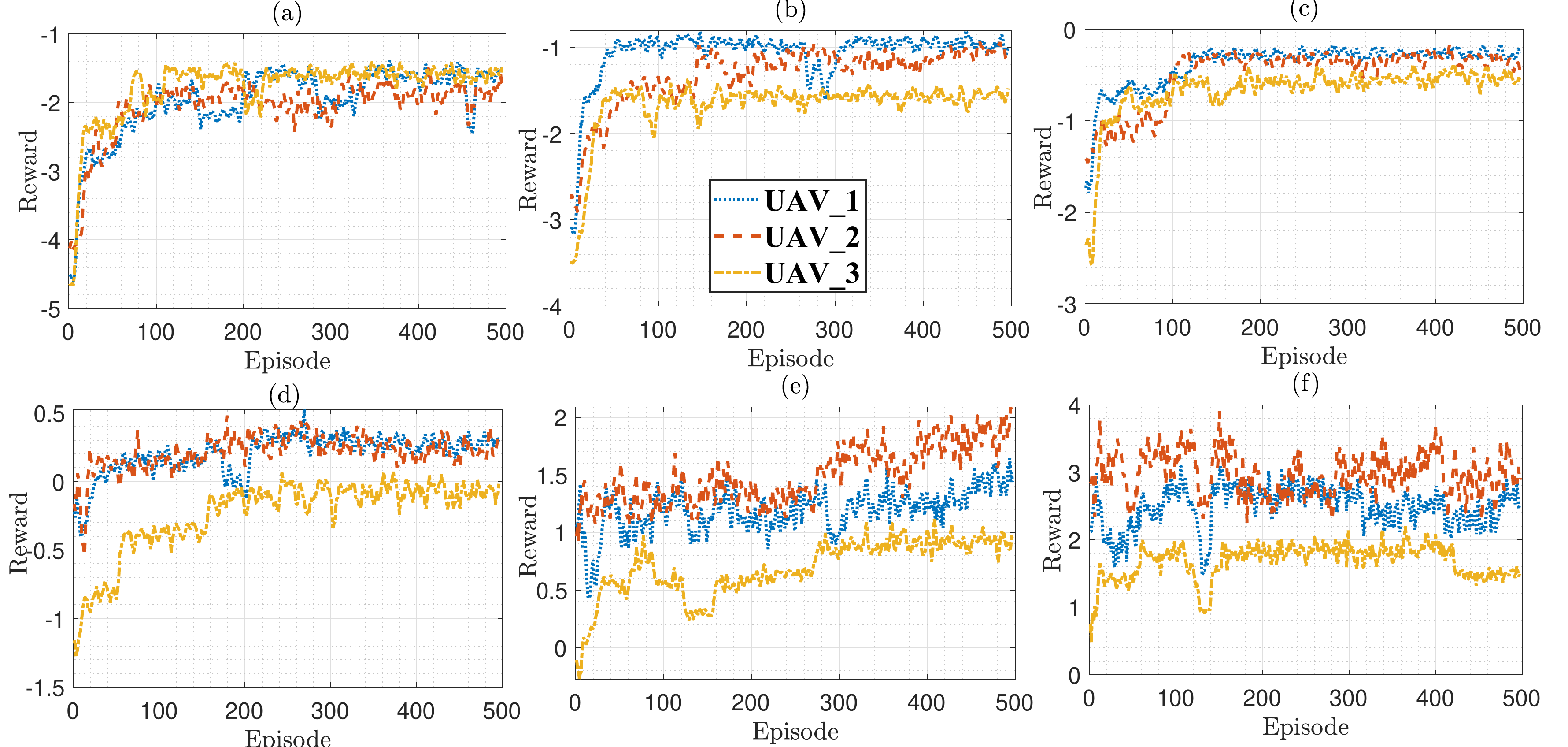}
	\caption{The rewards and how it converges: (a)  $\alpha=0.0$, (b)  $\alpha=0.2$, (c)  $\alpha=0.4$, (d)  $\alpha=0.6$, (e)  $\alpha=0.8$, and (f)  $\alpha=1.0$.}
	\label{fig:reward}
\end{figure*}
\section{simulation}
\label{section:simulation}
In this section, an evaluation of the performance of the proposed algorithm is presented via numerical results. The simulation settings are summarized in Table \ref{simulation}. For the main simulation, the number of UAVs is $3$ and $9$ users are moving on the ground and the environment is limited to a cylindrical with radius and height of $50$ m and  $100$ m, respectively. The time slot of the frame is considered to be $100$ ms and the slots are considered to be $1$ ms, in which the constant acceleration is specified in the frame, and in the slots the final speeds of each slot are considered as the initial speed of the next slot. The maximum speed for UAVs is 10 \si[per-mode=symbol]{\meter\per\second} in each direction, and the maximum acceleration is 2 \si[per-mode=symbol]{\meter\per\second^{2}}. All information about the neural network and the number of layers for each factor is given in Table \ref{simulation}. In the following, we will discuss all the figures obtained from the simulation. First, we will explain the baselines, and then we will examine the reward and the overall data rate of the network and the allocated powers, the effect of the minimum value of the data rate, the trajectory of the UAVs, and finally, the impact of constant velocity and constant acceleration on the performers. In addition, the source code of
the proposed modified MADDPG is available in \cite{4tg2-f112-21}.
\subsection{Solution Baselines}
The baselines include three solution methods as show in Fig. \ref{fig:baseline}, MADDPG, fully decentralized MADDPG, and DDPG, described below.
\begin{itemize}
	\item
	\textbf{MADDPG:} A standard method that operates on a multi-agent basis and has DDPG neural networks. The various agents interact with each other through a central server.
	\item 
	\textbf{Fully decentralized MADDPG:} In this solution method, agents work  independently of each other, have no contact, and only have their observations from the environment.
	\item 
	\textbf{DDPG:} The single agent interacts with the environment and has the whole environment as its neural network input.
\end{itemize}
\subsection{System model baselines}
\begin{itemize}
	\item
	\textbf{Non-CoMP:} A conventional system model without CoMP technology. Users with awful channels are deprived of receiving any assist from nearby UAVs.
	\item 
	\textbf{Constant velocity:} This type of motion is entirely unrealistic and is also a subset of accelerated motion in which the acceleration is zero, which is applied in UAV system model in previous works.
\end{itemize}
\subsection{Trade off between data rate and power consumption}
 Giving the weight factor $\alpha$, that changes the effect of each objective in the primary reward, we swipe this factor from 0 to 1 with step size of 0.2. Note that using this approach allows us to choose the priority between the power and the data rate  flexibly.
It can be seen in Fig. \ref{fig:reward}a that the goal is only to minimize the amount of power where only three constraints are considered. The required minimum data rate for each type of user and clipping the power between $0.1$ and $p^f_\text{max}$ helps stabilize the learning process. In this figure, the only goal of the agents is to meet the minimum rate for each type of user. In Fig. \ref{fig:reward}b, our reward is controlled by the power minimization instead of the data rate. In Fig. \ref{fig:reward}b, the sudden drop of the reward is due to the movement of users and handover, that the UAV is no longer able to track its users, and users switch among UAVs. Finally, in Fig. \ref{fig:reward}c, the effect of data rate exceeds the effect of power, and as we can observe in the figure, that rewards  converge to a positive value. However, the effect of power minimization also remains strong, but the priority is to maximize the data rate. 
In Fig. \ref{fig:reward}d, the impact of the data rate is increasing, and almost all agents policies are affected by the data rate of the whole system, and the sudden drop can be related to user movement and change in users assignment. In Fig. \ref{fig:reward}e, the impact of data rate increases, and at the beginning of learning, it can be observed that the agents try to increase the reward. However, the commitment to reduce power consumption will reduce reward in episodes afterward. It converges to another point
\begin{figure*}[ht!]
	\centering
	\includegraphics[width=1\linewidth]{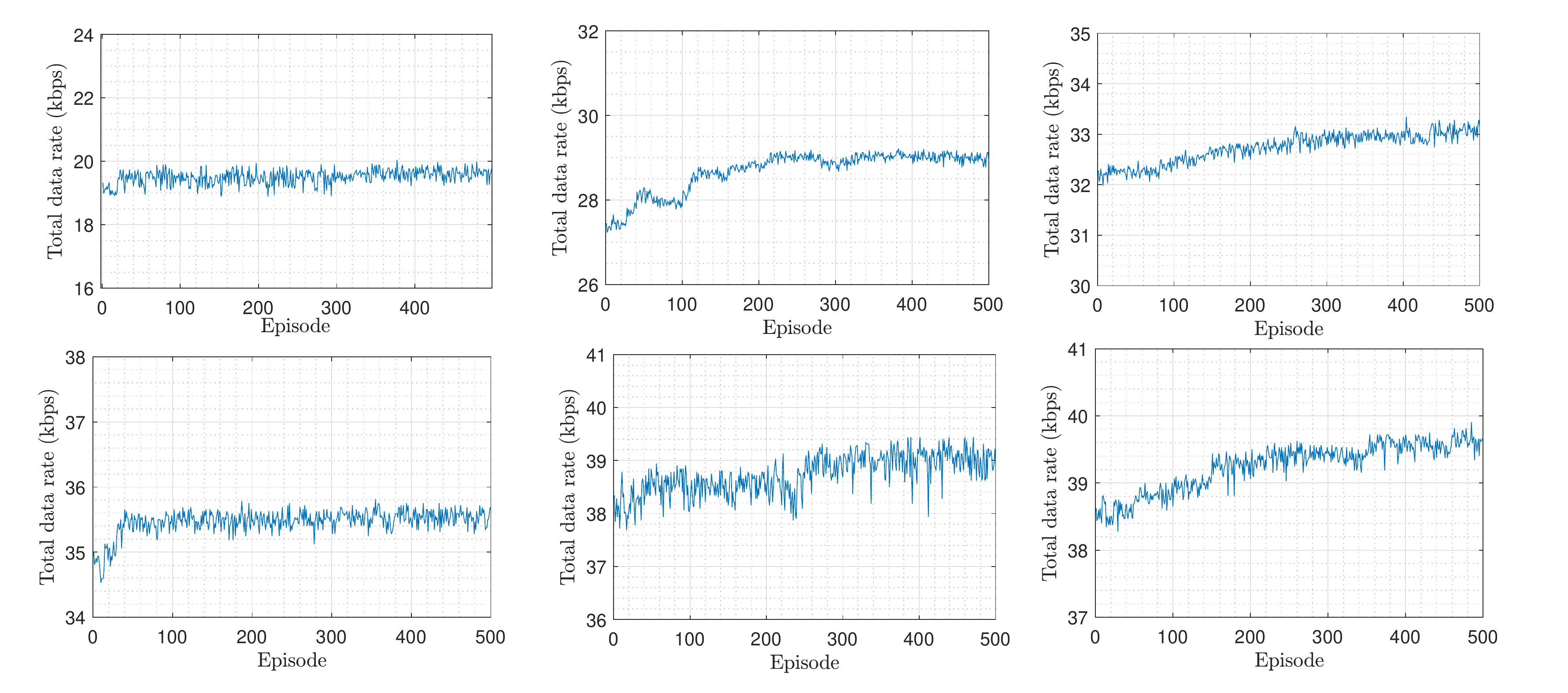}
	\caption{The data rate  and it convergence: (a)  $\alpha=0.0$, (b)  $\alpha=0.2$, (c)  $\alpha=0.4$, (d)  $\alpha=0.6$, (e)  $\alpha=0.8$, and (f)  $\alpha=1.0$.}
	\label{fig:totalrate}
\end{figure*} is due to the term of power minimization. Finally, in Fig. \ref{fig:reward}f, we see that the agents seek to maximize network data rates without  considering power consumption. \textcolor{black}{The fluctuation in Fig. \ref{fig:reward} can be attributed to the complexity of the system model and corresponding constraints, and the agents that  attempt to find the best case for the objective function.}
\begin{figure*}
	\centering
	\includegraphics[width=1\linewidth]{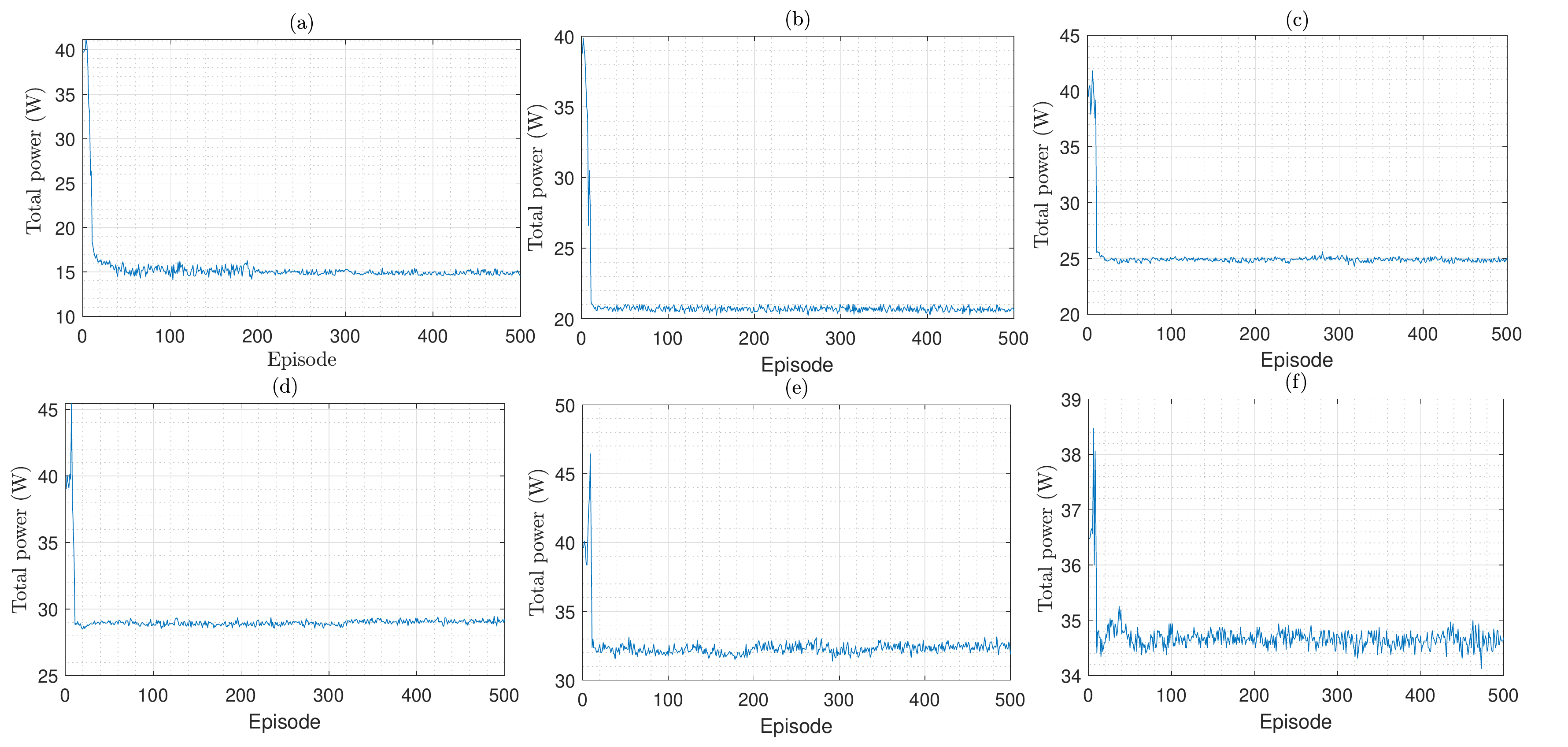}
	\caption{The total power consumption  and  it convergence: (a)  $\alpha=0.0$, (b)  $\alpha=0.2$, (c)  $\alpha=0.4$, (d)  $\alpha=0.6$, (e)  $\alpha=0.8$, and (f)  $\alpha=1.0$.}
	\label{fig:power-3}
\end{figure*}
\begin{figure}[h!]
	\centering
	\includegraphics[width=1\linewidth]{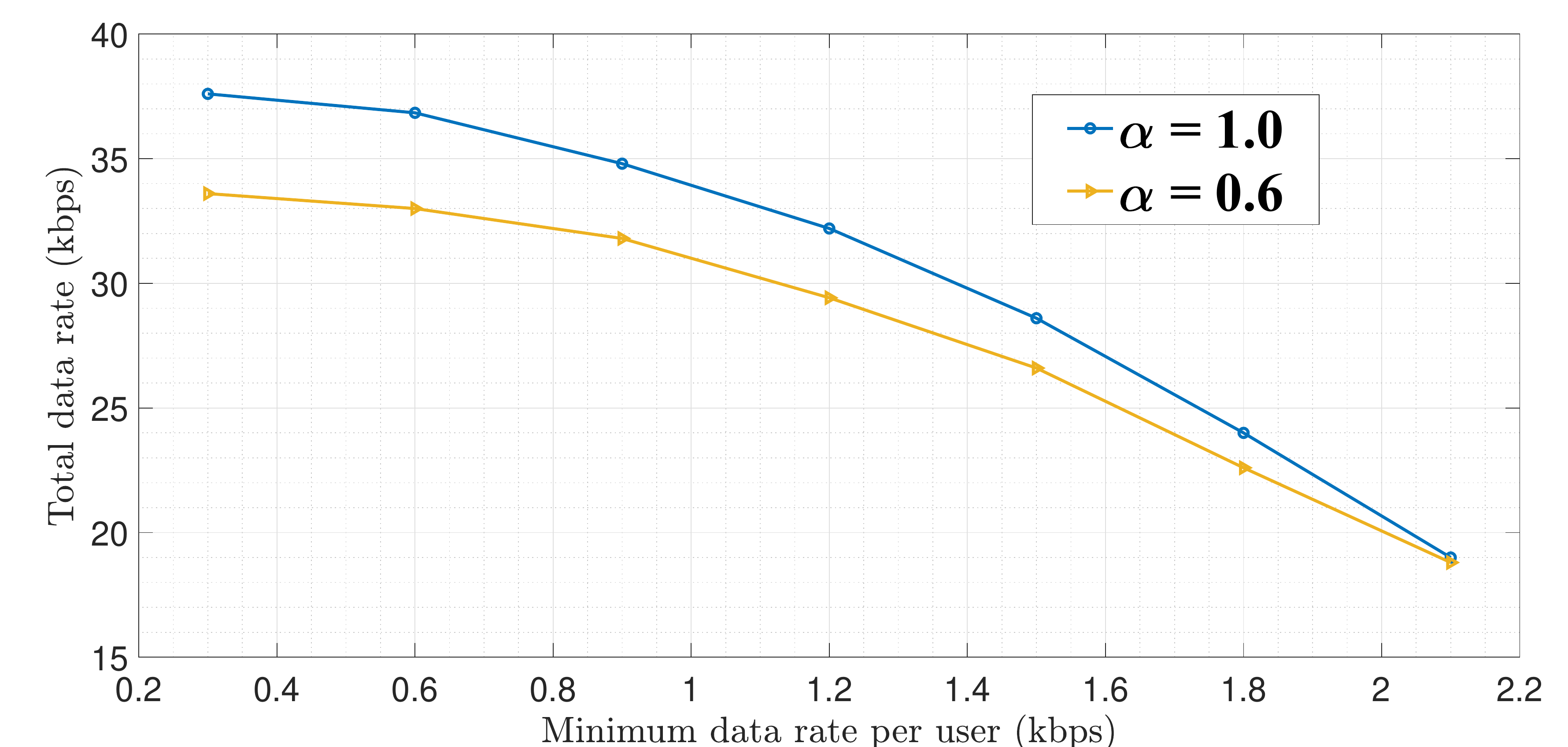}
	\caption{Effect of the minimum data rate on the total data rate.}
	\label{fig:min-data}
\end{figure}
\subsection{Data rate}
Here, we gather all the data rate figures in  Fig.  \ref{fig:totalrate} to compare with each other. In Fig.  \ref{fig:totalrate}a, the goal is only to minimize the total  power of system. In this case we do not have  the term to maximize the data rate, so it can be seen in the figure that the data rate changes instantly, and only three constraints are involved in this situation including minimum system data rate and power clipping between 0.1 and $p^f_\text{max}$ for a reason mentioned above. In Fig.  \ref{fig:totalrate}b, while increasing weighted factor $\alpha$, an attempt is initially made to increase the value of the data rate, but because the main priority is on power, the data rate decreases and converges to a lower number. Fig.  \ref{fig:totalrate}c shows the trade off between the data rate and power consumption, but as we can see in the figure, there are no obvious differences between the two terms of the reward. Furthermore, it seems that the two parts have neutralized each other's effect, although this is clearly shown in Fig. \ref{fig:totalrate}c, which shows the data rate. In Fig.  \ref{fig:totalrate}d, Fig.  \ref{fig:totalrate}e, and Fig.    \ref{fig:totalrate}f, as the weight of data rate ($\alpha$) in the reward increases, the priority changes in favor of increasing the data rate. However, due to the existence of a power control term, it is evident that the effect of the power minimization leads to decreasing the total data rate. The whole neural network strives to achieve the highest data rate.
\subsection{Power}
In Fig. \ref{fig:power-3}, we examine the transmission power of  UAVs in downlink communication. As mentioned in the previous sections, in Fig. \ref{fig:power-3}a, due to two constraints for the minimum data rate and clipping the power between 0.1 and $p^f_{\text{max}}$, the allocated power does not converge to zero, and agents try to establish the minimum data rate. By increasing the weight of  data rate ($\alpha$) in the reward, it can be noticed that the power consumption further increases to meet the constraints \eqref{rate-comp} and \eqref{rate-non}. In Fig. \ref{fig:power-3}e, it can be seen that there is a significant decrease in power consumption at the beginning of the run, which can be justified for the same reasons as above. In Fig. \ref{fig:power-3}f, we can observe that the system's whole purpose is to maximize the user data rate and meet the minimum requirements for the users, but the central server decides to  make a cooperation among the agents while reducing the interference. This central server also prevents the maximum power usage.
\subsection{Minimum data rate constraint}
The more we increase the minimum data rate constraint, the smaller the set of feasible answers gets. As illustrated in Fig. \ref{fig:min-data}, we continuously see a reduction in the data rate of the entire network by increasing the minimum data rate constraint, and this drop is more tangible in the higher numbers in the minimum data rate constraint.
\subsection{CoMP versus non-CoMP}
\begin{figure}
	\centering
	\includegraphics[width=1.05\linewidth]{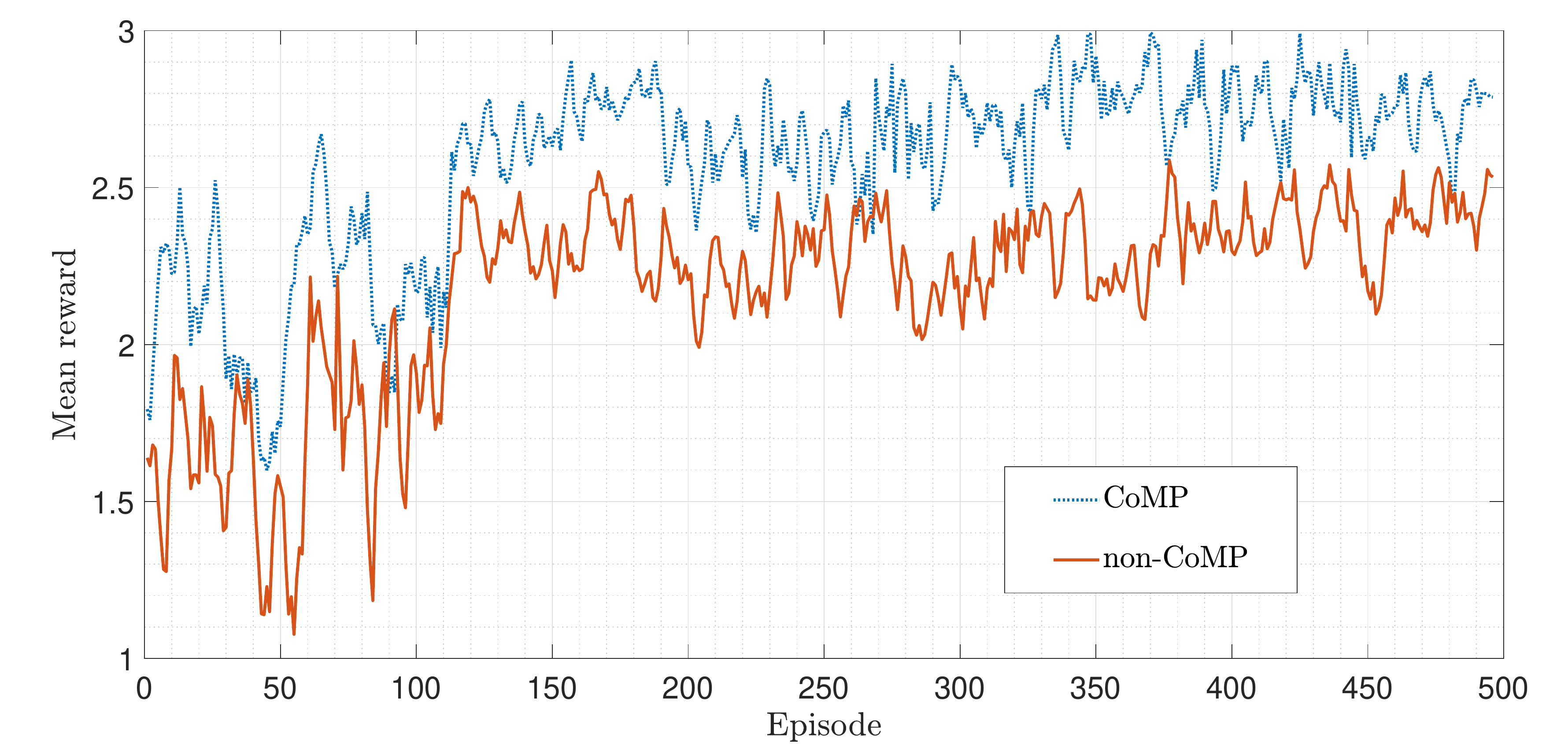}
	\caption{Comparing to sort of system model CoMP and non-CoMP $F=5$, $M=15$. }
	\label{fig:comp-enabled-vs-disabled}
\end{figure}

\begin{figure}
	\centering
	\includegraphics[width=1.05\linewidth]{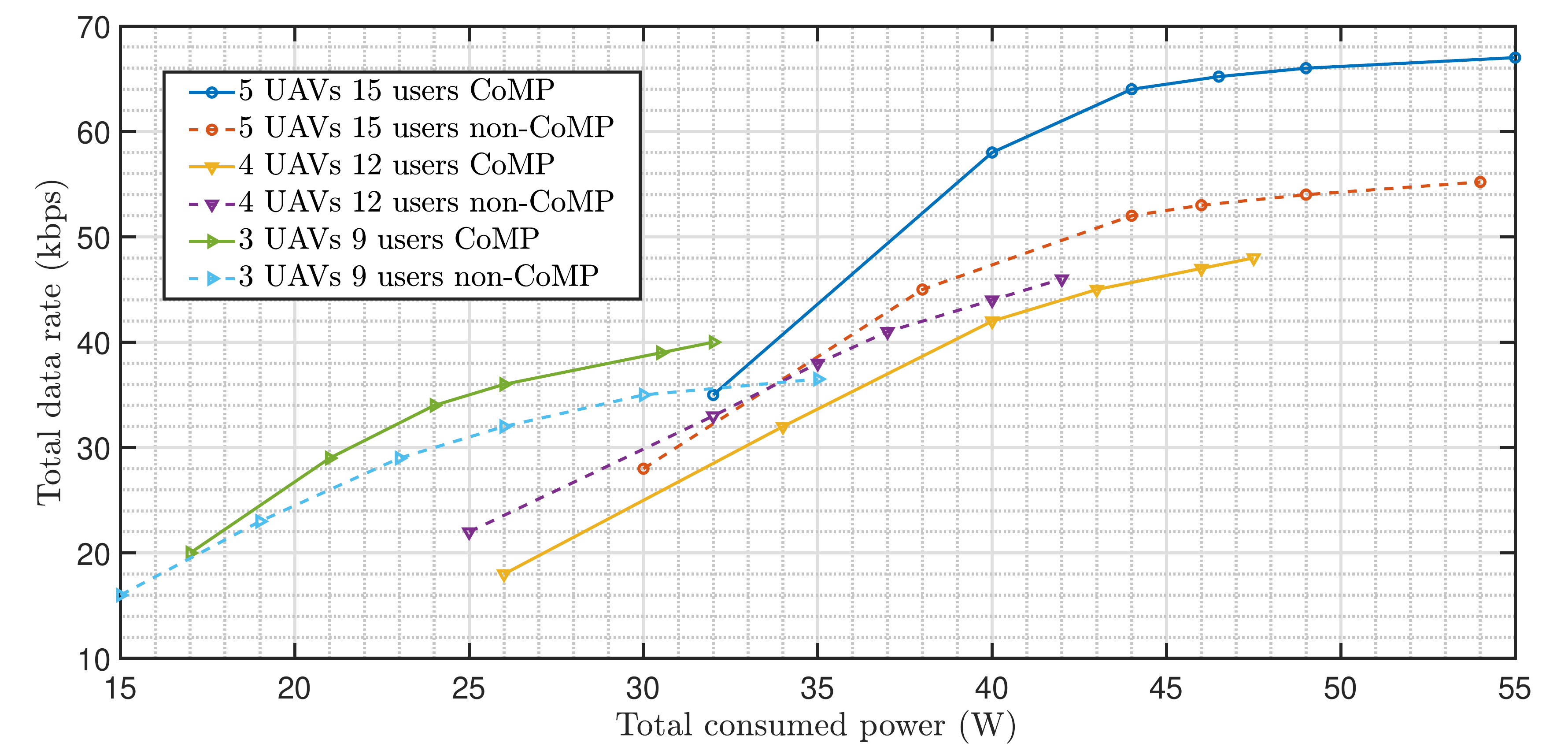}
	\caption{Data rate versus total power consumption CoMP and non-CoMP with varying number of the UAVs, the users, and values of  $\alpha$.}
	\label{fig:comperhensivecompenabled}
\end{figure}

\textcolor{black}{Fig. \ref{fig:comp-enabled-vs-disabled}  shows a complete run with  $F=5$ and $M=15$. On average, we observe  $12\%$ improvement of CoMP system in the reward compared with non-CoMP system. Fig. \ref{fig:comperhensivecompenabled} shows the variation of the number of UAVs and users, $\alpha$ value, and their impact on the data rate and the power. From this figure we can observe as the number of UAVs increases, the data rate increases. The power consumption of CoMP system is also significantly reduced in comparison with non-CoMP. The effect of CoMP is similar to the increasing of the number of users, i.e., using CoMP enhances the data rate.  CoMP users also reduce the power consumption of each UAV.}
\subsection{Constant acceleration versus constant velocity}
\begin{figure}
	\centering
	\includegraphics[width=1\linewidth]{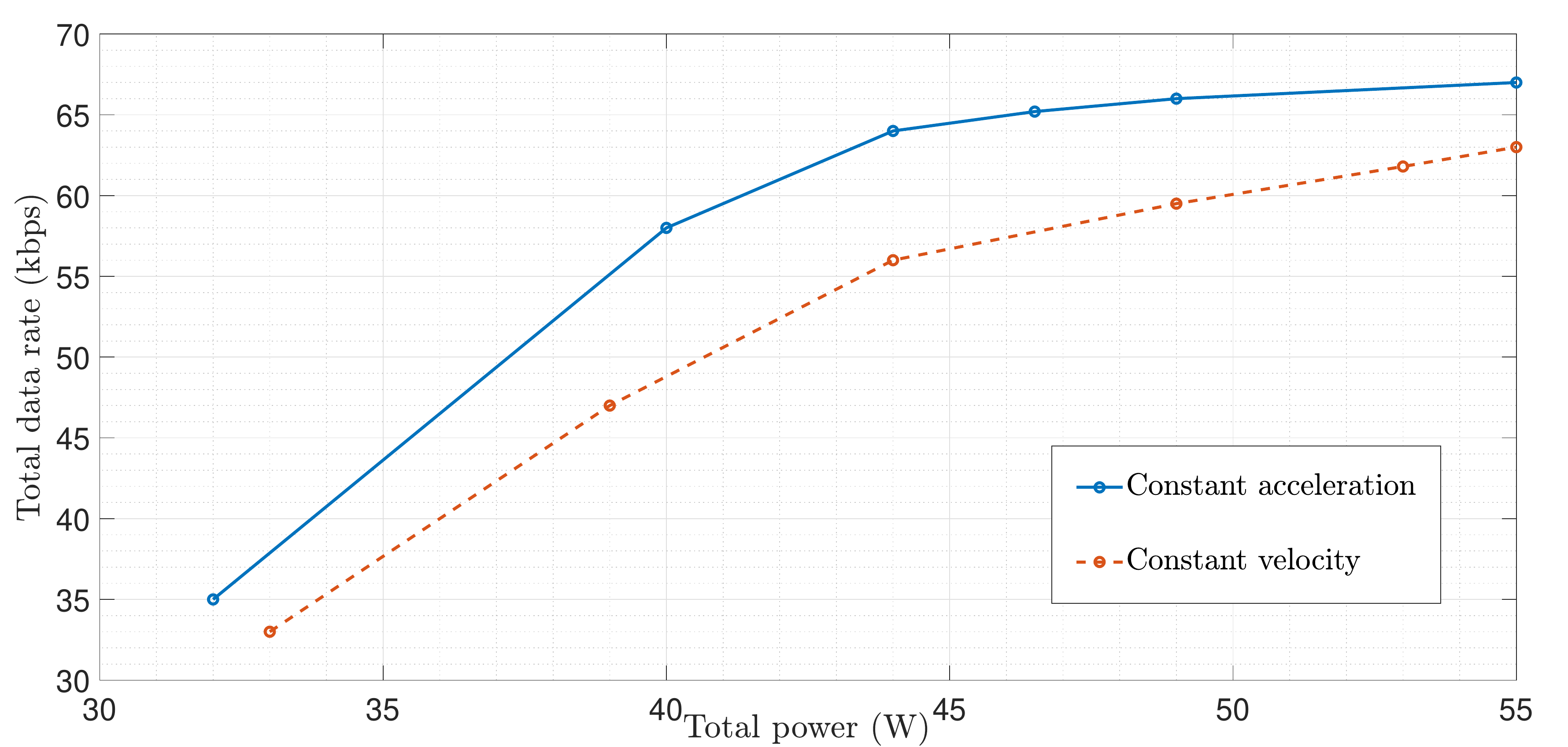}
	\caption{Constant acceleration versus constant velocity.}
	\label{fig:comprehensive-reward-rate-power}
\end{figure}

Here, we provide the same conditions regarding time and resource allocation to compare the two modes of motion with constant acceleration and motion with constant speed. From our simulation results considering the reward with weighted factor $ \alpha=0.8$, we perceive a better result through constant acceleration motion in terms of convergence value and convergence speed. For justifying this behavior, we can rely on the fact that moving at a constant acceleration makes UAVs more maneuverable and the ability to track the users on the ground better and faster as observed in Fig. \ref{fig:comprehensive-reward-rate-power}. \textcolor{black}{ Fig. \ref{fig:comprehensive-reward-rate-power}  shows that more data rates are allocated to the users, while, the amount of communication power consumption is significantly reduced. Also, the results are obtained for different values of $\alpha$.}
\section{Conclusion}
\label{section:Conclusion}
In this paper, we proposed a new frame structure to study the effect of constant acceleration on the CoMP UAV-enabled VLC networks. While giving two time frame, and using constant acceleration, which assists to improve the system efficiency, we solved this complex problem using novel machine learning and multi-agent method. Also, we presented a solution method based on our  system model. The results obtained from the simulation proved a better performance compared to other methods. The constant acceleration system  with  two time frame shows a better performance than the common system model including a constant velocity. As a future work, a novel RL method related to federated learning in swarm UAV networks can be investigated.
\bibliographystyle{ieeetr}
\bibliography{bbr}
\end{document}